\documentclass[sigconf,balance=true,hidelinks]{acmart}

\usepackage{hyperref}
\usepackage{algorithmic}
\usepackage[ruled,linesnumbered]{algorithm2e}
\usepackage{graphicx}
\usepackage{textcomp}
\usepackage{xcolor}
\usepackage{nccmath}
\usepackage{environ}
\usepackage{tcolorbox}
\usepackage{soul}
\usepackage{booktabs}
\usepackage{multirow}
\usepackage{array}
\usepackage{float}
\usepackage{makecell}
\usepackage{flushend}
\usepackage{comment}
\usepackage{enumitem}
\usepackage{epstopdf} 
\usepackage{caption}

\NewEnviron{myequation}{%
\begin{equation}
\scalebox{0.9}{$\BODY$}
\end{equation}
}

\newcolumntype{P}[1]{>{\centering\arraybackslash}p{#1}}

\setcopyright{none}

\begin{document}

\title{A Large-scale Study of Security Vulnerability Support on Developer Q\&A Websites}

\author{Triet Huynh Minh Le}
\affiliation{\institution{The University of Adelaide}
\city{Adelaide}
\country{Australia}}
\email{triet.h.le@adelaide.edu.au}

\author{Roland Croft}
\affiliation{\institution{The University of Adelaide}
\city{Adelaide}
\country{Australia}}
\affiliation{Cyber Security Cooperative Research Centre
\country{Australia}}
\email{roland.croft@adelaide.edu.au}

\author{David Hin}
\affiliation{\institution{The University of Adelaide}
\city{Adelaide}
\country{Australia}}
\affiliation{Cyber Security Cooperative Research Centre
\country{Australia}}
\email{david.hin@adelaide.edu.au}

\author{M. Ali Babar}
\affiliation{\institution{The University of Adelaide}
\city{Adelaide}
\country{Australia}}
\affiliation{Cyber Security Cooperative Research Centre
\country{Australia}}
\email{ali.babar@adelaide.edu.au}

\begin{abstract}
\textbf{Context}: Security Vulnerabilities (SVs) pose many serious threats to software systems. Developers usually seek solutions to addressing these SVs on developer Question and Answer (Q\&A) websites. However, there is still little known about on-going SV-specific discussions on different developer Q\&A sites. \textbf{Objective}: We present a large-scale empirical study to understand developers' SV discussions and how these discussions are being supported by Q\&A sites. \textbf{Method}: We first curate 71,329 SV posts from two large Q\&A sites, namely Stack Overflow (SO) and Security StackExchange (SSE). We then use topic modeling to uncover the topics of SV-related discussions and analyze the popularity, difficulty, and level of expertise for each topic. We also perform a qualitative analysis to identify the types of solutions to SV-related questions. \textbf{Results}: We identify 13 main SV discussion topics on Q\&A sites. Many topics do not follow the distributions and trends in expert-based security sources such as Common Weakness Enumeration (CWE) and Open Web Application Security Project (OWASP). We also discover that SV discussions attract more experts to answer than many other domains, but some difficult SV topics (e.g., Vulnerability Scanning Tools) still receive quite limited support from experts. Moreover, we identify seven key types of answers given to SV questions on Q\&A sites, in which SO often provides code and instructions, while SSE usually gives experience-based advice and explanations.
\textbf{Conclusion}: Our findings provide support for researchers and practitioners to effectively acquire, share and leverage SV knowledge on Q\&A sites.
\end{abstract}



\maketitle

\section{Introduction}
It is important to constantly track and resolve Security Vulnerabilities (SVs) to ensure the availability, confidentiality and integrity of software systems~\mbox{\cite{ghaffarian2017software}}. Developers can seek information for resolving SVs from sources verified by security experts such as Common Weakness Enumeration (CWE), National Vulnerability Database (NVD) and Open Web Application Security Project (OWASP).
However, these expert-based SV sources do not provide any mechanisms for developers to promptly ask and answer questions about issues in implementing/understanding the reported SV solutions/concepts.
On the other hand, developer Questions and Answer (Q\&A) websites contain a plethora of such SV-related discussions. Stack Overflow (SO) and Security StackExchange\footnote{https://security.stackexchange.com/} (SSE) contain some of the largest number of SV-related discussions among developer Q\&A sites, with contributions from millions of users~\mbox{\cite{le2020puminer}}.

The literature has analyzed different aspects of discussions on Q\&A sites, but there is still no investigation of how SO and SSE are supporting SV-related discussions. Specifically, the main concepts~\mbox{\cite{yang2016security}}, the top languages/technologies and user demographics~\mbox{\cite{bayati2016information}}, as well as user perceptions and interactions~\mbox{\cite{lopez2019anatomy}} of general security discussions on SO have been studied. However, from our analysis (see section~\mbox{\ref{subsec:post_collection}}), only about 20\% of the available SV posts on SO were investigated in the previous studies, limiting a thorough understanding about SV topics (developers' concerns when tackling SVs in practice) on Q\&A sites. Moreover, the prior studies only focused on SO, and little insight has been given into the support of SV discussions on different Q\&A sites that can affect the choice of a suitable site (e.g., SO vs. SSE) to discuss certain SV topics.

To fill these gaps, we conduct a large-scale empirical study using 71,329 SV posts on SO and SSE. Specifically, we use Latent Dirichlet Allocation (LDA)~\mbox{\cite{blei2003latent}} topic modeling and qualitative analysis to answer the following four Research Questions (RQs) that measure the support of Q\&A sites for different SV discussion topics:

\noindent \textbf{RQ1}: What are SV discussion topics on Q\&A sites?

\noindent\textbf{RQ2}: What are the popular and difficult SV topics?

\noindent \textbf{RQ3}: What is the level of expertise for supporting SV questions?

\noindent \textbf{RQ4}: What types of answers are given to SV questions?

\noindent Our findings to these RQs can help raise developers' awareness of common SVs and enable them to seek solutions to such SVs more effectively on Q\&A sites. We also identify the areas to which experts can contribute to assist the secure software engineering community. Furthermore, we release a large dataset of SV discussions on Q\&A sites for replication and future work at~\mbox{\cite{reproduction_package_ease2021}}.

\section{Related Work}
\label{sec:related_work}
\subsection{Topic Modeling on Q\&A Websites}

Q\&A websites such as SO and SSE contain a large number of discussion posts. LDA~\cite{blei2003latent} has been frequently used to extract the taxonomy/topics of various software-related domains from such posts. In 2014, a seminal work of Barua et al.~\cite{barua2014developers} discovered the topics of all SO posts. They also found that LDA could find more consistent topics than the tags on SO. Many subsequent studies have leveraged LDA to investigate discussions of specific domains, such as general security~\cite{yang2016security}, concurrent computing~\cite{ahmed2018concurrency}, mobile computing~\cite{rosen2016mobile}, big data~\cite{bagherzadeh2019going}, machine learning~\cite{bangash2019developers} and deep learning~\cite{han2020programmers}. Among the aforementioned studies, Yang et al.~\mbox{\cite{yang2016security}} is the closest to our work. However, our work is still fundamentally different from this previous study. Despite sharing a similar security context to Yang et al.~\mbox{\cite{yang2016security}}, we focus specifically on the flaws of security implementation/features since exploitation of such flaws can disclose user's data and interrupt system operations. For example, authentication is an important security requirement to enforce different levels of access/permission in a system, but credentials of an authentication process may be vulnerable to timing or brute-force attacks. Moreover, we consider the content of both questions and answers of SV posts on two Q\&A sites (SO and SSE) rather than just questions on SO as in~\mbox{\cite{yang2016security}}. This gives more in-depth insights into how different Q\&A sites are supporting on-going SV discussions. Detailed discussion on these differences is given in section~\mbox{\ref{subsec:vs_literature}}.

\vspace{-4pt}

\subsection{Security Vulnerability Analytics Using Open Sources}

SV analytics have long been of interest to researchers. Frei et al.~\cite{frei2006large} were among the first to study the SV life cycle using open-source security advisories. Shahzad et al.~\cite{shahzad2012large} then conducted a large-scale study on the characteristics (e.g., risk metrics, exploitation, affected vendors and products) of reported SVs on NVD. There is another active research trend to build prediction models to analyze SVs. Bozorgi et al.~\cite{bozorgi2010beyond} used Support Vector Machine to predict the probability and time-to-exploit of SVs. There have been many follow-up studies since then on developing/improving Machine Learning (e.g.,~\cite{le2019automated,spanos2018multi}) and Deep Learning (e.g.,~\cite{han2017learning,sahin2019conceptual}) models to determine various properties of SVs using expert-based SV sources (e.g., CWE and NVD). A recent study~\cite{horawalavithana2019mentions} leveraged security mentions on social media (i.e., Twitter and Reddit) to forecast the SV-related activities on GitHub. Unlike the above studies, we focus on SV analytics on developer Q\&A sites. Several studies (e.g.,~\mbox{\cite{meng2018secure,rahman2019snakes}}) analyzed SVs of different programming languages using code snippets on SO. Contrary to these studies, we do not limit our investigation to any specific programming language, and we consider every type of SV-related posts, not just the ones with code snippets.

\vspace{-4pt}

\section{Research Method}
\label{sec:case_study}

\subsection{Research Questions}
\label{subsec:rq_method}

We investigated four Research Questions (RQs) to provide a thorough picture of the support of Q\&A websites for Security Vulnerability (SV) related discussions.
To answer these RQs, we retrieved 71,329 SV posts from a general Q\&A website (Stack Overflow (SO)) and a security-centric one (Security StackExchange (SSE)) using both the tags and content of posts, as described in section~\ref{subsec:post_collection}.

\noindent \textbf{RQ1: What are SV discussion topics on Q\&A sites?}\label{rq1_method}

\noindent \underline{\textit{Motivation}}: To provide fine-grained information about the support of SO and SSE for different types of SV discussions, we first needed to identify the taxonomy of commonly discussed SV topics in RQ1. Our taxonomy does not aim to replace the existing ones provided by experts (e.g., CWE or OWASP), but rather helps to highlight the important aspects of SVs from developers' perspective.

\noindent \underline{\textit{Method}}: Following the standard practice in~\mbox{\cite{barua2014developers,yang2016security,rosen2016mobile,ahmed2018concurrency,bagherzadeh2019going,bangash2019developers,han2020programmers}}, RQ1 used Latent Dirichlet Allocation (LDA) \cite{blei2003latent} topic modeling technique (see section~\ref{subsec:lda}) to select SV discussion topics based on the titles, questions and answers of SV posts on both SO and SSE. LDA is commonly used since it can produce topic distribution (assigning multiple topics with varying relevance) for a post, providing more flexibility/scalability than manual coding. We also used the topic share metric~\cite{barua2014developers} in Eq.~\eqref{eq:share} to compute the proportion ($\text{share}_{i}$) of each SV topic and their trends over time.

\setlength{\abovedisplayskip}{2pt}
\setlength{\belowdisplayskip}{2pt}

\begin{myequation}\label{eq:share}
\text{share}_{i}=\frac{1}{N}\sum\limits_{p\,\in \,D}{\text{LDA}(p,\,\,{{\text{T}}_{i}})}
\end{myequation}

\noindent where $p$, $D$ and $N$ are a single SV post, the list of all SV posts and the number of such posts, respectively; $\text{T}_{i}$ is the $i^{\text{th}}$ topic and LDA is the trained LDA model.

\noindent \textbf{RQ2: What are the popular and difficult SV topics?}\label{rq2_method}

\noindent \underline{\textit{Motivation}}: After the SV topics were identified, RQ2 identified the popular and difficult topics on Q\&A websites. The results of RQ2 can aid the selection of a suitable (i.e., more popular and less difficult) Q\&A site for respective SV topics.

\noindent \underline{\textit{Method}}: To quantify the topic popularity, we used four metrics from~\cite{rosen2016mobile,yang2016security, bagherzadeh2019going,ahmed2018concurrency}, namely the average values of (\textit{i}) views, (\textit{ii}) scores (upvotes minus downvotes), (\textit{iii}) favorites and (\textit{iv}) comments. Intuitively, a more popular topic would attract more attention (views), interest (scores/favorites) and activities (comments) per post from users. We also obtained the geometric mean of the popularity metrics to produce a more consistent result across different topics. Geometric mean was used instead of arithmetic mean here since the metrics could have different units/scales. To measure the topic difficulty, we used the three metrics from~\cite{rosen2016mobile,yang2016security, bagherzadeh2019going,ahmed2018concurrency}: (\textit{i}) percentage of getting accepted answers, (\textit{ii}) median time (hours) to receive an accepted answer since posted, and (\textit{iii}) average ratio of answers to views. A more difficult topic would, on average, have a lower number of accepted answers and ratio of answers to views, but a higher amount of time to obtain accepted answers. To achieve this, we took reciprocals of the difficulty metrics (\textit{i}) and (\textit{iii}) so that a more difficulty topic had a higher geometric mean of the metrics.

\noindent \textbf{RQ3: What is the level of expertise to answer SV questions?}\label{rq3_method}

\noindent \underline{\textit{Motivation}}: RQ3 checked the expertise level available on Q\&A websites to answer SV questions, especially the ones of difficult topics. The findings of RQ3 can shed light on which topic may require more attention from experts. Note that experts here are users who frequently contribute helpful (accepted) answers/knowledge.

\noindent \underline{\textit{Method}}: We measured both user's general and specific expertise for SV topics on Q\&A sites. For the general expertise, we leveraged the commonly used metric, the reputation points~\mbox{\cite{hanrahan2012modeling,meng2018secure,rahman2019snakes}}, of users who got accepted answers since reputation is gained through one's active participation and appreciation from the Q\&A community in different topics. Higher reputation received for a topic usually implies that the questions of that topic are of more interest to experts.
Similar to~\mbox{\cite{meng2018secure}}, we did not normalize the reputation by user's participation time since reputation may not increase linearly, e.g., due to users leaving the sites.\footnote{http://varianceexplained.org/r/are\_users\_quitting/} However, reputation is not specific to any topic; thus it does not reflect whether a user is experienced with a topic. Hence, we represented developers' specific expertise with the SV content in their answers on Q\&A sites. This was inspired by Dey et al.'s findings that developers' expertise/knowledge could be expressed through their generated content~\mbox{\cite{dey2020representation}}. We determined a user's expertise in SV topics using the topic distribution generated by LDA applied to the concatenation of all answers to SV questions given by that user. The specific expertise of an SV topic (see Eq.~\mbox{\eqref{eq:spec_exp}}) was then the total correlation between LDA outputs of the current topic in SV questions and the specific expertise of users who got the respective accepted answers. The correlation of LDA values could reveal the knowledge (SV topics) commonly used to answer questions of a certain (SV) topic~\mbox{\cite{barua2014developers}}.

\begin{myequation}\label{eq:spec_exp}
\begin{aligned}
Specific\_Expertis{{e}_{i}}=\sum\limits_{p\in D}{\text{LDA}(Q(p),\,\,{{\text{T}}_{i}})\odot \text{LDA}(K({{U}_{\text{Accept}\text{.}}}))} \\ 
K({{U}_{\text{Accept}\text{.}}})=A_{{{U}_{\text{Accept}\text{.}}}}^{1}+A_{{{U}_{\text{Accept}\text{.}}}}^{2}+...+A_{{{U}_{\text{Accept}\text{.}}}}^{k}(k=\left| {{A}_{{{U}_{\text{Accept}\text{.}}}}} \right|)
\end{aligned}
\end{myequation}

\noindent where $D$ is the list SV posts and $\text{T}_{i}$ is the $i^{\text{th}}$ topic, while $Q(p)$ and $K({{U}_{\text{Accept}\text{.}}})$ are the question content and SV knowledge of the user ${U}_{\text{Accept}\text{.}}$ who gave the accepted answer of the post $p$, respectively. $\odot$ is the topic-wise multiplication. $\left| {{A}_{{{U}_{\text{Accept}\text{.}}}}} \right|$ is all SV-related answers given by user ${U}_{\text{Accept}\text{.}}$. Note that we only considered posts with accepted answers to make it consistent with the general expertise.

\noindent Specifically, for each question, we first extracted the user that gave the accepted answer (${U}_{\text{Accept}\text{.}}$). We then gathered all answers, not necessarily accepted, of that user in SV posts ($\left| {{A}_{{{U}_{\text{Accept}\text{.}}}}} \right|$). Such answer list was the SV knowledge of ${U}_{\text{Accept}\text{.}}$ ($K({{U}_{\text{Accept}\text{.}}})$).
Finally, we computed the LDA topic-wise correlation between the topic $\text{T}_{i}$ in the current SV question ($\text{LDA}(Q(p),\,\,{{\text{T}}_{i}})$) and the user knowledge ($\text{LDA}(K({{U}_{\text{Accept}\text{.}}}))$) to determine the specific expertise for post $p$.

\noindent \textbf{RQ4: What types of answers are given to SV questions?}\label{rq4_method}

\noindent \underline{\textit{Motivation}}: RQ4 extended RQ2 in terms of the solution types given if an SV question is satisfactorily answered. We do not aim to provide solutions for every single SV. Rather, we analyze and compare the types of support for different SV topics on SO and SSE, which can guide developers to a suitable site depending on their needs (e.g., looking for certain artefacts). To the best of our knowledge, we are the first to study answer types of SVs on Q\&A sites.

\noindent \underline{\textit{Method}}: We employed an open coding procedure~\cite{seaman1999qualitative} to inductively identify answer types. LDA is \textit{not suitable} for this purpose since it relies on word co-occurrences to determine categories. In contrast, the same type of solutions may not share any similar words. In RQ4, we only considered the posts with accepted answer to ensure the high quality and relevance of the answers. We then used stratified sampling to randomly select 385 posts (95\% confidence level with 5\% margin error~\cite{cochran2007sampling}) each from SO and SSE to categorize the answer types. Stratification ensured the proportion of each topic was maintained.
Following~\mbox{\cite{chen2020comprehensive}}, two of the authors first conducted a pilot study to assign initial codes to 30\% of the selected posts and grouped similar codes into answer types. For example, the accepted answers of SO posts 32603582 (PostgreSQL code), 20763476 (MySQL code) and 12437165 (Android/Java code) were grouped into \textit{Code Sample} category. Similarly to~\mbox{\cite{treude2011programmers}}, we also allowed one post to have more than one answer type.
Two same authors then independently assigned the identified categories to the remaining 70\% of the posts. The Kappa inter-rater score ($\kappa$)~\cite{mchugh2012interrater} was 0.801 (strong agreement), showing the reliability of our coding. The third author involved to discuss and resolve the disagreements. We also correlated the answer types with the question types on Q\&A sites~\cite{treude2011programmers}.

\begin{figure*}[t]
    \centering
    {%
    \setlength{\fboxsep}{3.0pt}%
    \setlength{\fboxrule}{0.5pt}%
    \fbox{\includegraphics[width=0.97\textwidth,keepaspectratio]{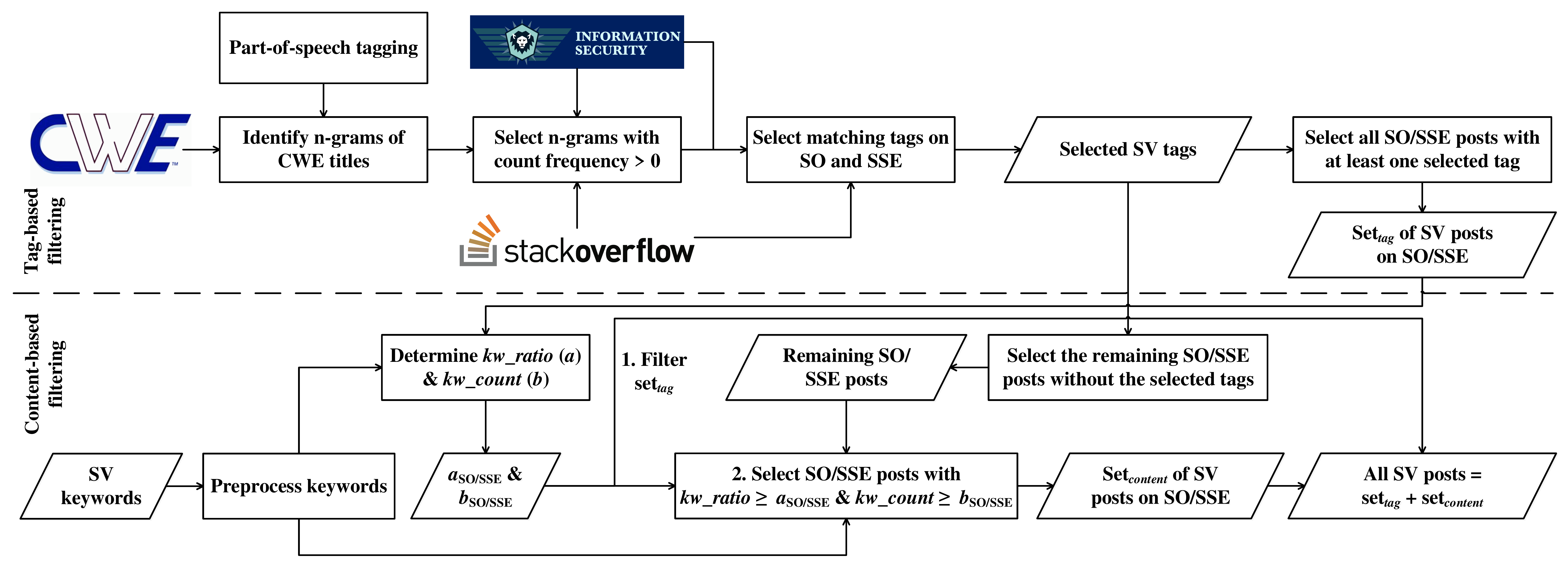}
    }}
    \caption{Workflow of retrieving posts related to SV on Q\&A websites using tag-based and content-based filtering heuristics.}
    \label{fig:workflow}
\end{figure*}

\subsection{Security Vulnerability Post Collection}
\label{subsec:post_collection}

To study the support of Q\&A sites for SV discussions, we proposed a workflow (see Fig.~\ref{fig:workflow}) to obtain, to the best of our knowledge, the largest and most contemporary set of SV posts on both SO and SSE. We used \textit{tag-based} and \textit{content-based filtering} to retrieve SV posts based on their tags and content of other parts (i.e., title, body and answers), respectively. We considered a post to be related to SV when it mainly discussed a security flaw and/or exploitation/testing/fixing of such flaw to compromise a software system (e.g., SO post 29098142\footnote{stackoverflow.com/questions/29098142 (postid: 29098142). SSE format is security.stackexchange.com/questions/postid. Posts in our paper follow these formats.}). A post was not SV-related if it just asked how to implement/use a security feature (e.g., SO post 685855) without any explicit mention of a flaw. All the tags, keywords and posts collected were released at~\cite{reproduction_package_ease2021}.

\noindent \textbf{Tag-based filtering}.
We had a \textit{vulnerability} tag on SSE but not on SO to obtain SV-related posts, and a \textit{security} tag on SO used by~\mbox{\cite{yang2016security}} was too coarse-grained for the SV domain. Many posts with \textit{security} tag did not explicitly mention SV (e.g., SO post 65983245 about privacy or SO post 66066267 about how to obtain security-relevant commits). Therefore, we used Common Weakness Enumeration (CWE), which contains various SV-related terms, to define relevant SV tags. However, the full CWE titles were usually long and uncommonly used in Q\&A discussions. For example, the fully-qualified CWE name of SQL-injection (CWE-89) is ``\emph{Improper Neutralization of Special Elements used in an SQL Command (`SQL Injection')}'', which appeared only nine times on SO and SSE. Therefore, we needed to extract shorter and more common terms from the full CWE titles. We adopted Part-of-Speech (POS) tagging for this purpose, in which we only considered consecutive (n-grams of) verbs, nouns and adjectives since most of them conveyed the main meaning of a title. For instance, we obtained the following 2-grams for CWE-89: \textit{improper neutralization}, \textit{special elements}, \textit{elements used}, \textit{sql command}, \textit{sql injection}. We obtained 2,591 n-gram (1 $\leq$ n $\leq$ 3) terms that appeared at least once on either SO or SSE. To ensure the relevance of these terms, we manually removed the irrelevant terms without any specific SV context (e.g., \textit{special elements}, \textit{elements used} and \textit{sql command} in the above example). We found 60 and 63 SV-related tags on SO and SSE that matched the above n-grams, respectively. We then obtained the initial $\text{set}_{tag}$ of SV posts that had at least one of these selected tags.

\noindent \textbf{Content-based filtering}. As recommended by some recent studies~\cite{haque2020challenges,le2020puminer}, tag-based filtering was not sufficient for selecting posts due to wrong tags (e.g., non SV-post 38539393 on SO with \textit{stack-overflow} tag) or general tags (e.g., SV post 15029849 on SO with only \textit{php} tag). Therefore, as depicted in Fig.~\ref{fig:workflow}, we customized content-based filtering, which was based on keyword matching, to refine the $\text{set}_{tag}$ obtained from the tag-based filtering step and select missing SV posts that were not associated with SV tags. First, we presented the up-to-date list of 643 SV keywords for matching~\cite{reproduction_package_ease2021}. These keywords were preprocessed with stemming and augmented with American/British spellings, space/hyphen to better handle various types of (mis-)spellings/plurality. For instance, we considered the following variants: \emph{input(-)sanitization/sanit/sanitisation/sanitis} for ``\emph{input sanitization}''. Similar to~\cite{haque2020challenges,le2020puminer}, we also performed exact matching for three-character keywords and subword matching for longer ones to reduce false positives. Subsequently, for each $\text{set}_{tag}$ (SO and SSE) obtained in the tag-based filtering step, we computed two content-based metrics (see Eq.~\eqref{eq:content_heuristics})~\cite{haque2020challenges,le2020puminer}: $kw\_count$ and $kw\_ratio$, denoting the count and appearance proportion of SV keywords in a post, respectively. $Kw\_count$ ensured diverse SV-related content in a post, while $kw\_ratio$ increased the confidence that these relevant words did not appear by chance.

\begin{myequation}\label{eq:content_heuristics}
kw\_coun{{t}_{p}}=\left| SV\_KW{{s}_{p}} \right|\,,\,\,\,kw\_rati{{o}_{p}}=\frac{\left| SV\_KW{{s}_{p}} \right|}{\left| Word{{s}_{p}} \right|}
\end{myequation}

\noindent where $\left| SV\_KW{{s}_{p}} \right|$ and $\left| Word{{s}_{p}} \right|$ are the numbers of SV keywords and total number of words in post $p$, respectively.

\noindent Based on the post content and human inspection, the thresholds $a_{\text{SO/SSE}}$ and $b_{\text{SO/SSE}}$ for filtering $\text{set}_{tag}$ (step 1) as well as selecting extra posts based on their content (step 2) were found, as given in Table~\ref{tab:content_thresholds}. Using these thresholds, we obtained $\text{set}_{tag}$ and $\text{set}_{content}$ of SV posts on SO and SSE, respectively, as shown in Fig.~\ref{fig:workflow}.

\begin{table}[t]
\fontsize{8}{9}\selectfont
  \centering
  \caption{Content-based thresholds ($a_{\text{SO/SSE}}$ \& $b_{\text{SO/SSE}}$) for the two steps of the content-based filtering as shown in Fig.~\ref{fig:workflow}.}
  \vspace{-6pt}
    \begin{tabular}{|c|P{1.3cm}|P{1.3cm}|P{1.3cm}|P{1.3cm}|}
    \hline
    \multirowcell{2}{\textbf{Thres-}\\ \textbf{hold}} & \multicolumn{2}{c|}{\textbf{Stack Overflow (SO)}} & \multicolumn{2}{c|}{\makecell{\textbf{Security}\\ \textbf{StackExchange (SSE)}}} \\
    \cline{2-5}
    & \textbf{Step 1} & \textbf{Step 2} & \textbf{Step 1} & \textbf{Step 2} \\
    \hline
    \textbf{a} & 1 & 3 & 2 & 3 \\
    \hline
    \textbf{b} & 0.011 & 0.017 & 0.017 & 0.025 \\
    \hline
    \end{tabular}%
  \label{tab:content_thresholds}%
\end{table}%

\noindent \textbf{SV datasets and validation}. As of June 30, 2020, we retrieved 20,062,329 and 58,912 posts from SO and SSE, respectively, using Stack Exchange Data Explorer. We then applied the tag-based and content-based filtering steps in Fig.~\ref{fig:workflow} and obtained \textit{71,329} SV posts (see Table~\ref{tab:sv_post_stats}) in total including 55,883 and 15,436 ones for $\text{set}_{tag}$ and $\text{set}_{content}$, respectively. We manually validated four different sets of SV posts, i.e., $\text{set}_{tag}$ and $\text{set}_{content}$ for SO and SSE, respectively. Specifically, we randomly sampled 385 posts (significant size~\cite{cochran2007sampling}) in each set for two authors to examine independently.

\begin{table}[t]
\fontsize{8}{9}\selectfont
  \centering
  \caption{The obtained SV posts using our tag-based and content-based filtering heuristics.}
  \vspace{-6pt}
    \begin{tabular}{|l|c|c|c|}
    \hline
     & \makecell{\textbf{Stack Over-}\\ \textbf{flow (SO)}} & \makecell{\textbf{Security Stack-}\\ \textbf{Exchange (SSE)}} & \textbf{SO + SSE}\\
    \hline
    \textbf{Set\textsubscript{\emph{tag}}} & 46,212 & 9,677 & 55,889 \\
    \hline
    \textbf{Set\textsubscript{\emph{content}}} & 12,660 & 2,780 & 15,440 \\
    \hline
    \textbf{Total} & 58,872 & 12,457 & 71,329 \\
    \hline
    \end{tabular}%
  \label{tab:sv_post_stats}%
\end{table}%

For $\text{set}_{tag}$, we disagreed on 7/770 cases and only two posts were not related to SV. The main issue was still the incorrect tag assignment (e.g., SSE post 175264 was about dll injection but tagged with \textit{malware}\footnote{This post was short yet contained many SV keywords (e.g., ``\textit{injection}'' and ``\textit{hijack}''), resulting in high $kw\_count$ and $kw\_ratio$ of the content-based filtering.}), though this issue had been significantly reduced by the content-based filtering. For $\text{set}_{content}$, the relevance of the posts was very high as there was no discrepant case.

Our SV dataset was only 20\% overlapping with the existing security dataset~\cite{yang2016security}, implying that there were significant differences in the nature of the two studies. Note that we followed the settings in~\cite{yang2016security} to retrieve the updated security posts from the same SO data we used in our study. We also reported the top tags of SV posts (see Table~\ref{tab:top_tags}) and compared them with the ones of security posts~\cite{yang2016security} and a subset of all the posts containing an equal number of posts to the SV posts on SO and SSE. SV posts were associated with many SV-related tags (e.g., \emph{memory-leaks}, \emph{malware}, \emph{segmentation-fault}, \emph{xss}, \emph{exploit} and \emph{penetration-test}). Conversely, security posts were tagged with general terms that may not explicitly discuss security flaws such as \emph{encryption}, \emph{authentication} and \emph{passwords}. The tags of general posts were mostly programming languages on SO and general security terms on SSE. These findings highlight the importance of obtaining SV-specific posts instead of reusing the security posts to study the support of Q\&A sites for SV-related discussions.

\begin{table}[t]
\fontsize{8}{9}\selectfont
  \centering
  \caption{Top-5 tags of SV, security and general posts on SO and SSE (in parentheses).}
  \vspace{-6pt}
    \begin{tabular}{|c|c|c|c|}
    \hline
    \textbf{No.} & \multicolumn{1}{c|}{\textbf{SV posts}} & \multicolumn{1}{c|}{\textbf{Security posts}} & \multicolumn{1}{c|}{\textbf{General posts}} \\
    \hline
    1     & \makecell{memory-leaks\\ (malware)} & \makecell{security\\ (encryption)} & \makecell{javascript\\ (encryption)} \\
    \hline
    2     & \makecell{segmentation-\\fault (web-appli-\\cation)} & encryption (tls) & java (tls) \\
    \hline
    3     & php (xss) & \makecell{php (authentication)} & \makecell{python\\(authentication)} \\
    \hline
    4     & c (exploit) & java (passwords) & \makecell{c\#\\(passwords)} \\
    \hline
    5     & \makecell{security\\(penetration-test)} & \makecell{cryptography\\(web-application)} & \makecell{php\\ (certificates)} \\
    \hline
    \end{tabular}%
  \label{tab:top_tags}%
  \vspace{-6pt}
\end{table}%


\subsection{Topic Modeling with LDA}
\label{subsec:lda}

Following the common practice of the existing work (e.g.,~\cite{yang2016security,barua2014developers,bagherzadeh2019going}), we extracted the topics of the identified SV-related posts on both SO and SSE using Latent Dirichlet Allocation (LDA)~\cite{blei2003latent}.

\noindent \textbf{Preprocessing of SV posts}. The first step was to preprocess posts to avoid noise in the text, helping a topic model to identify more relevant words of each topic.
Following the previous practices (e.g.,~\cite{yang2016security,ahmed2018concurrency}), we first removed the HTML tags and code snippets in each post as these elements were not informative for topic modeling.
We also converted the text to lowercase, removed punctuations, and then eliminated stop words and performed stemming (reducing a word to its root form) to avoid irrelevant and multi-form words.

\noindent \textbf{Topic modeling with LDA}. We applied LDA to the title, question body and all answers of each Q\&A post.
Regarding the number of topics ($k$) of LDA, we examined an inclusive range from 2 to 80, with an increment of one topic at a time. As suggested in~\mbox{\cite{barua2014developers,rosen2016mobile,ahmed2018concurrency,bagherzadeh2019going}}, alongside $k$, we also tried different values of $\alpha$ (1/$k$ or 50/$k$) and $\beta$ (0.01 or same as $\alpha$) hyperparameters to optimize the performance of LDA. $\alpha$ controls the sparsity of the topic-distribution per post and $\beta$ determines the sparsity of the word-distribution per topic. For each tuple of ($k$, $\alpha$ and $\beta$), we ran LDA with 1,000 iterations, then evaluated the coherence metric~\mbox{\cite{roder2015exploring}} of the identified topics. Coherence metric has been recommended by many previous studies (e.g.,~\cite{abdellatif2020challenges,zahedi2020mining}) to select the optimal number of LDA topics since it usually highly correlates with human understandability. Topic coherence is the average correlation between pairs of words that appear in the same topic. The higher value of the coherence metric means the more coherent content of the posts within a same topic.
To avoid insignificant topics like~\cite{barua2014developers}, we only considered topics with a probability of at least 0.1 in a post.
We manually read the top-20 most frequent words and 15 random posts of each topic per site (SO/SSE) obtained by the trained LDA models to label the name of that topic as done in~\cite{bagherzadeh2019going,ahmed2018concurrency}.
The LDA model with most relevant set of topics would be used for answering the four RQs.

\begin{table}[t]
\fontsize{8}{9}\selectfont
  \centering
  \caption{SV topics on SO and SSE identified by LDA along with their proportions and trends over time. \textbf{Notes}: The topic proportions on SSE are in parenthesis. The trends of SO are the top solid sparklines, while the trends of SSE are the bottom dashed sparklines. Unit of proportion: \%.}
  \vspace{-6pt}
    \begin{tabular}{|c|c|c|}
    \hline
    \textbf{Topic Name} & \textbf{Proportion} & \textbf{Trend} \\
    \hline
    Malwares (T1) & 1.39 (8.18) & \parbox[c]{0.7cm}{\includegraphics[width=0.8cm,keepaspectratio]{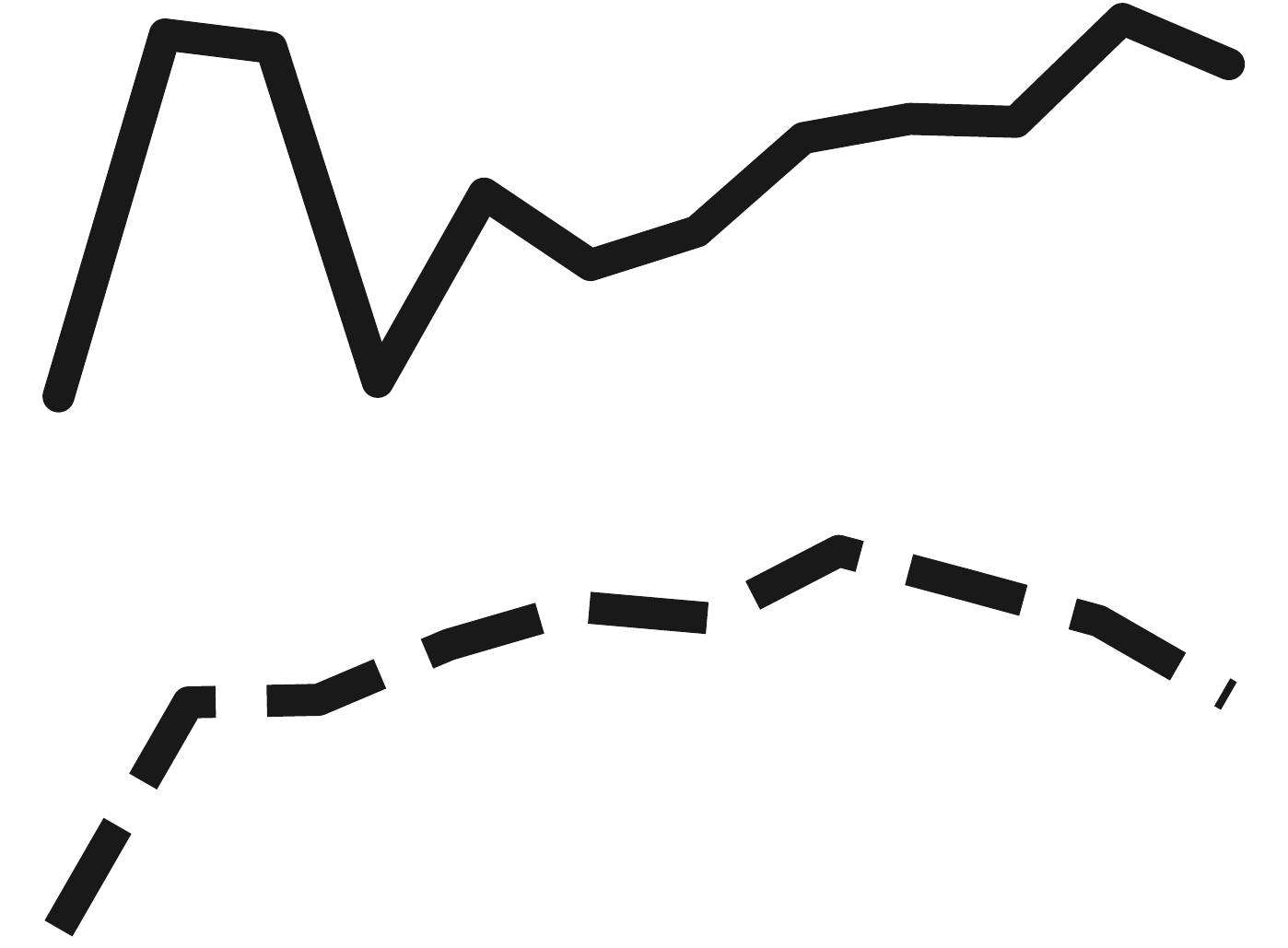}} \\
    \hline
    SQL Injection (T2) & 11.0 (4.17) & \parbox[c]{0.7cm}{\includegraphics[width=0.8cm,keepaspectratio]{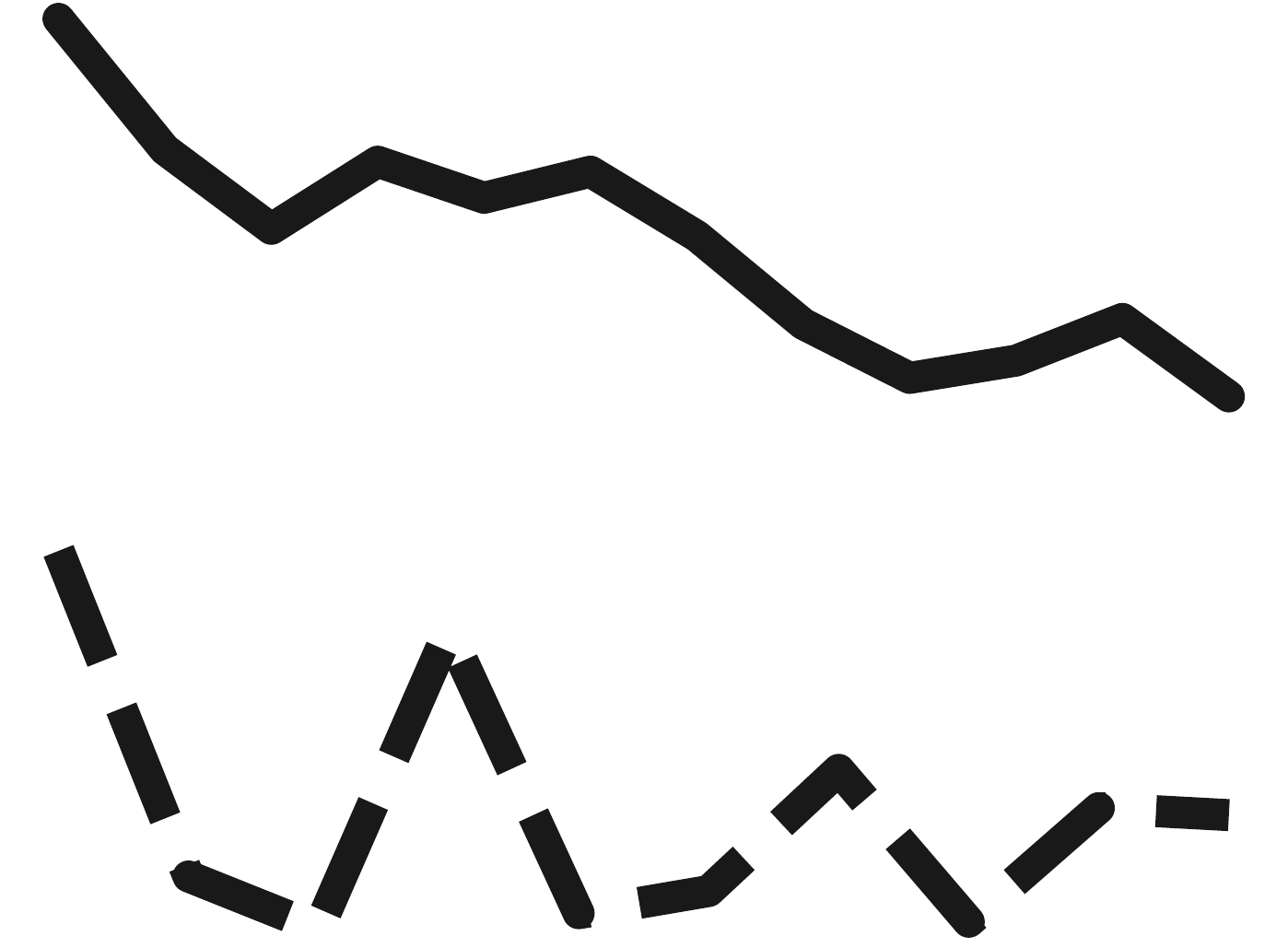}} \\
    \hline
    Vulnerability Scanning Tools (T3) & 5.42 (3.15) & \parbox[c]{0.7cm}{\includegraphics[width=0.8cm,keepaspectratio]{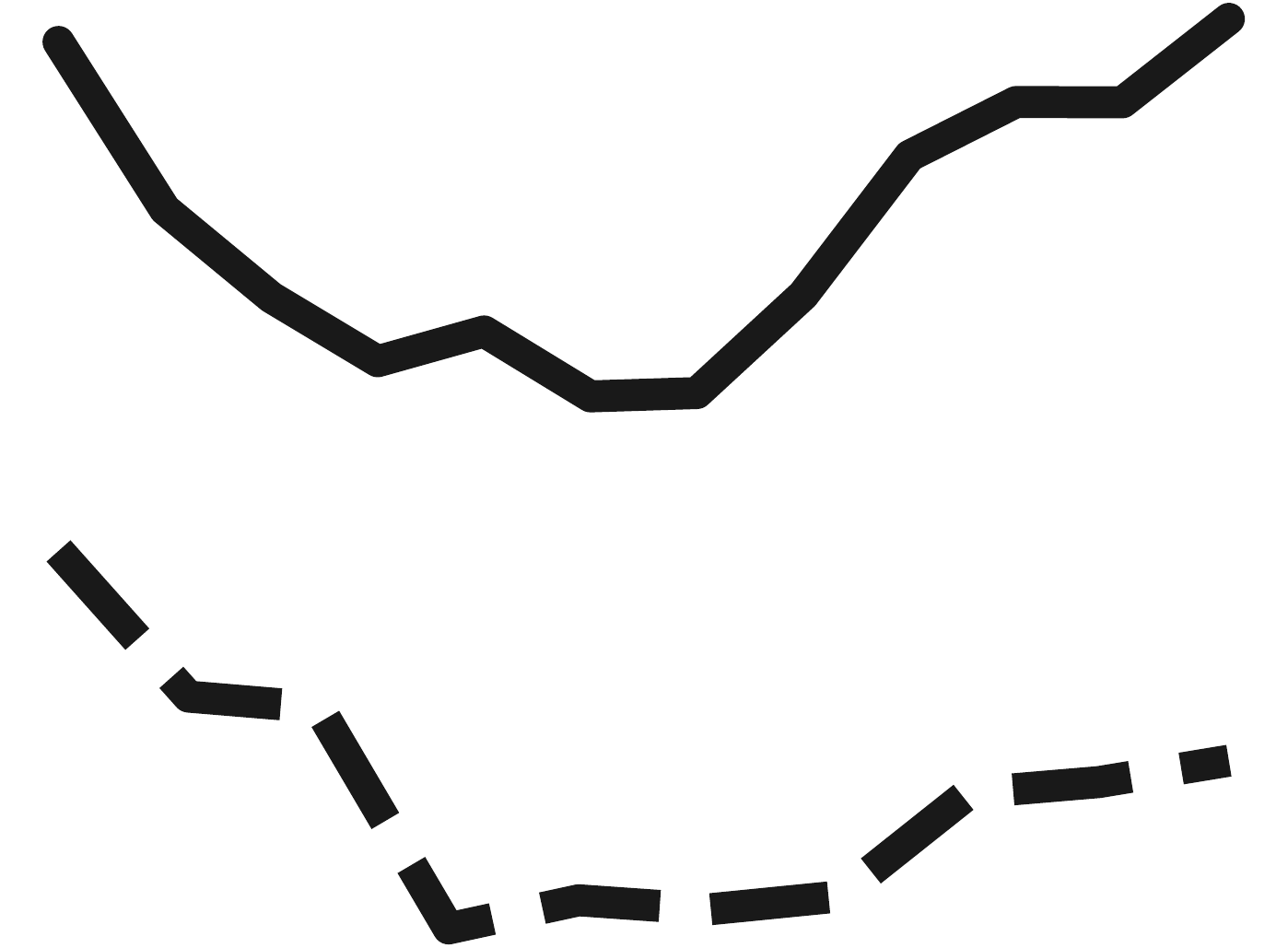}}\\
    \hline
    Cross-site Request Forgery (CSRF) (T4) & 9.49 (5.09) & \parbox[c]{0.7cm}{\includegraphics[width=0.8cm,keepaspectratio]{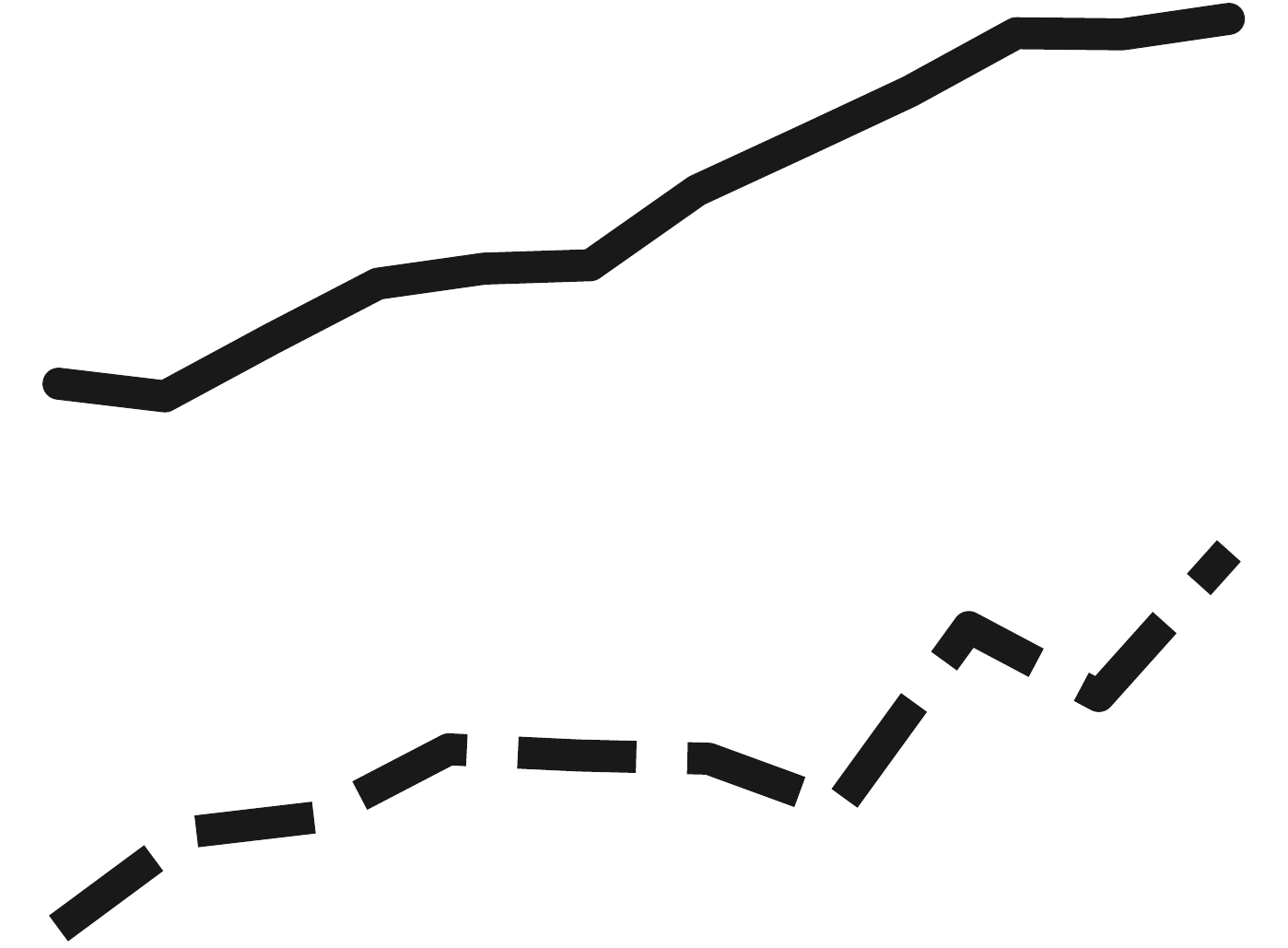}} \\
    \hline
    File-related Vulnerabilities (T5) & 2.88 (3.24) & \parbox[c]{0.7cm}{\includegraphics[width=0.8cm,keepaspectratio]{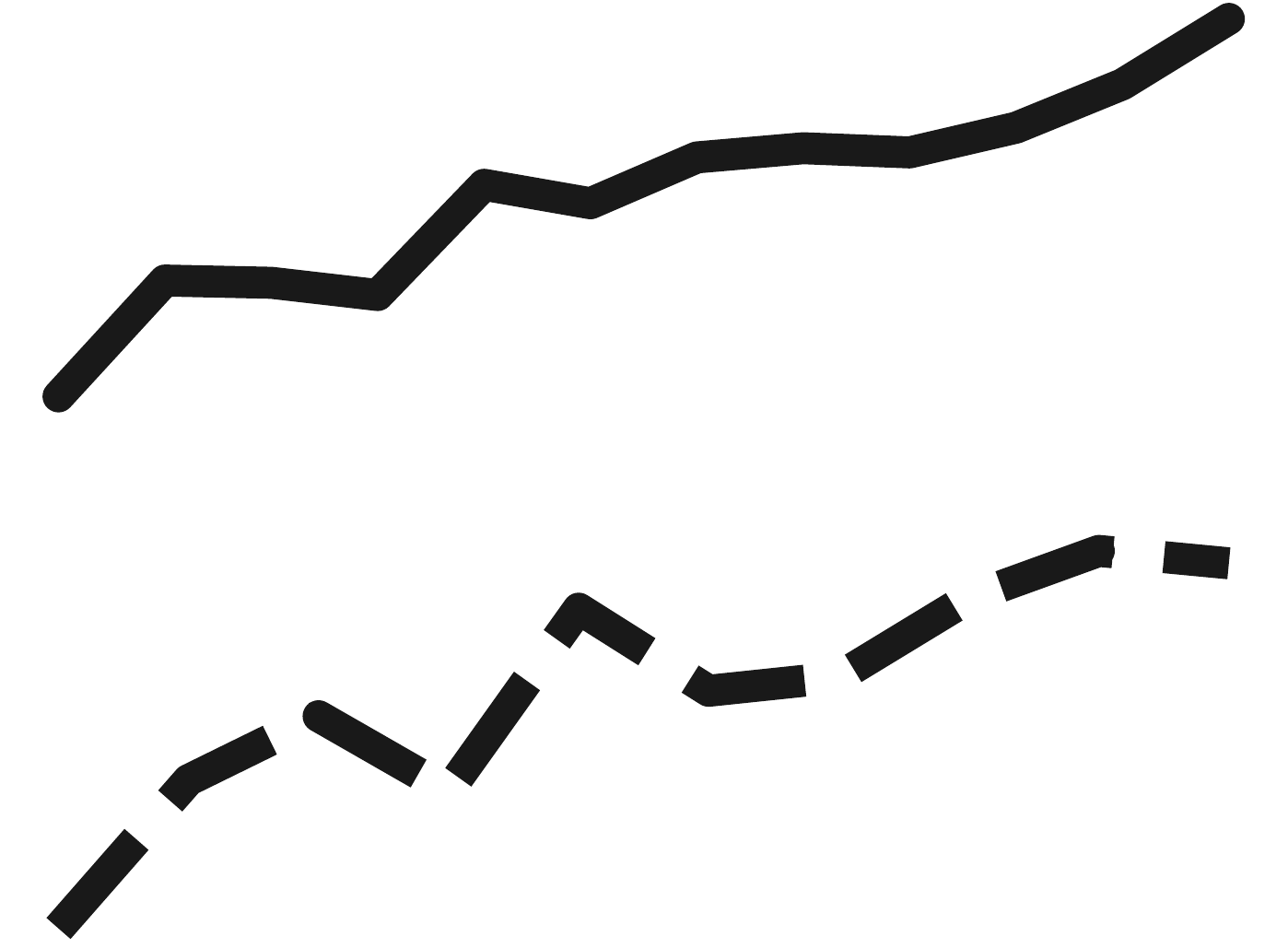}} \\
    \hline
    Synchronization Errors (T6) & 3.79 (0.47) & \parbox[c]{0.7cm}{\includegraphics[width=0.8cm,keepaspectratio]{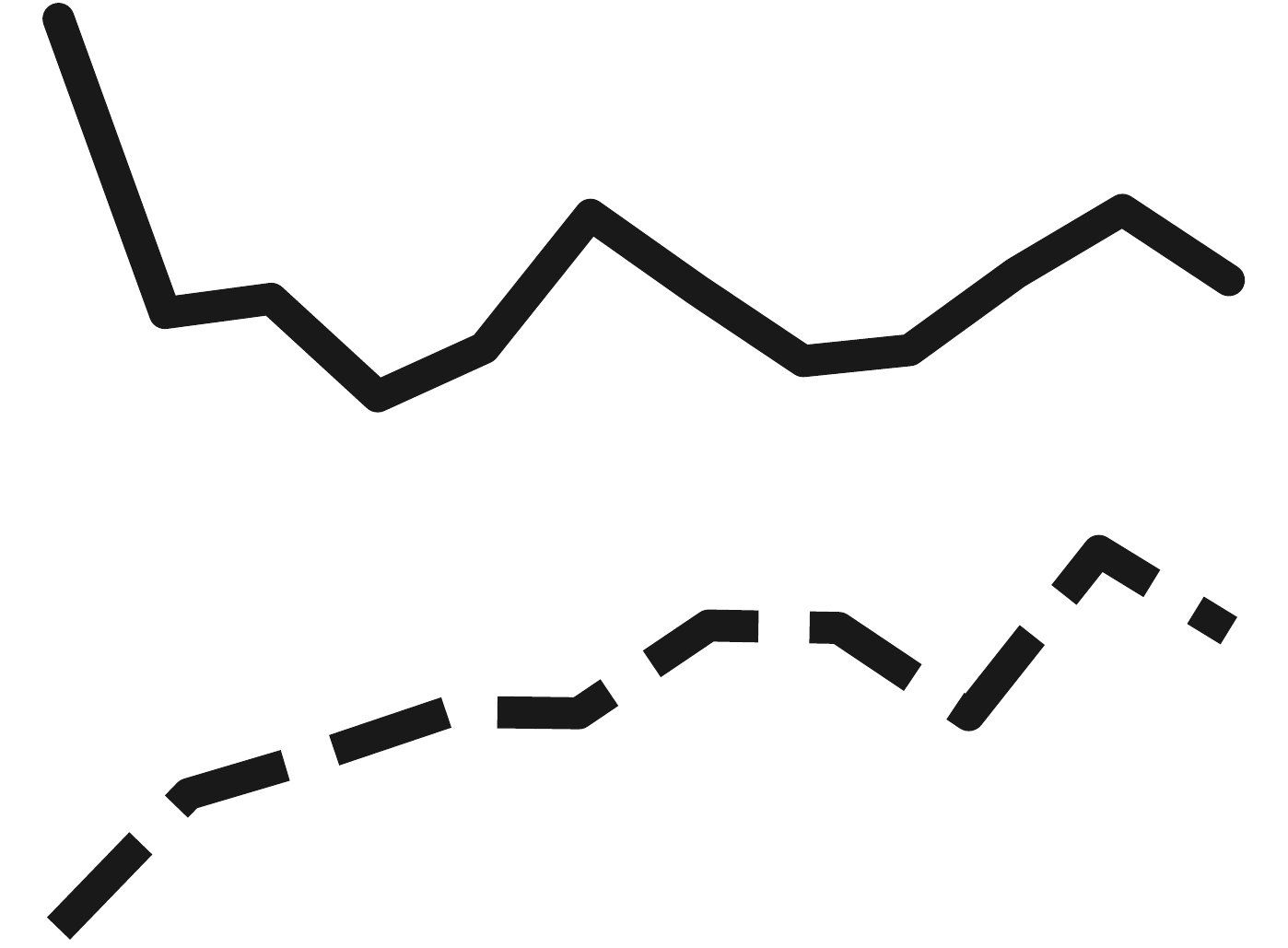}} \\
    \hline
    Encryption Errors (T7) & 1.82 (7.81) & \parbox[c]{0.7cm}{\includegraphics[width=0.8cm,keepaspectratio]{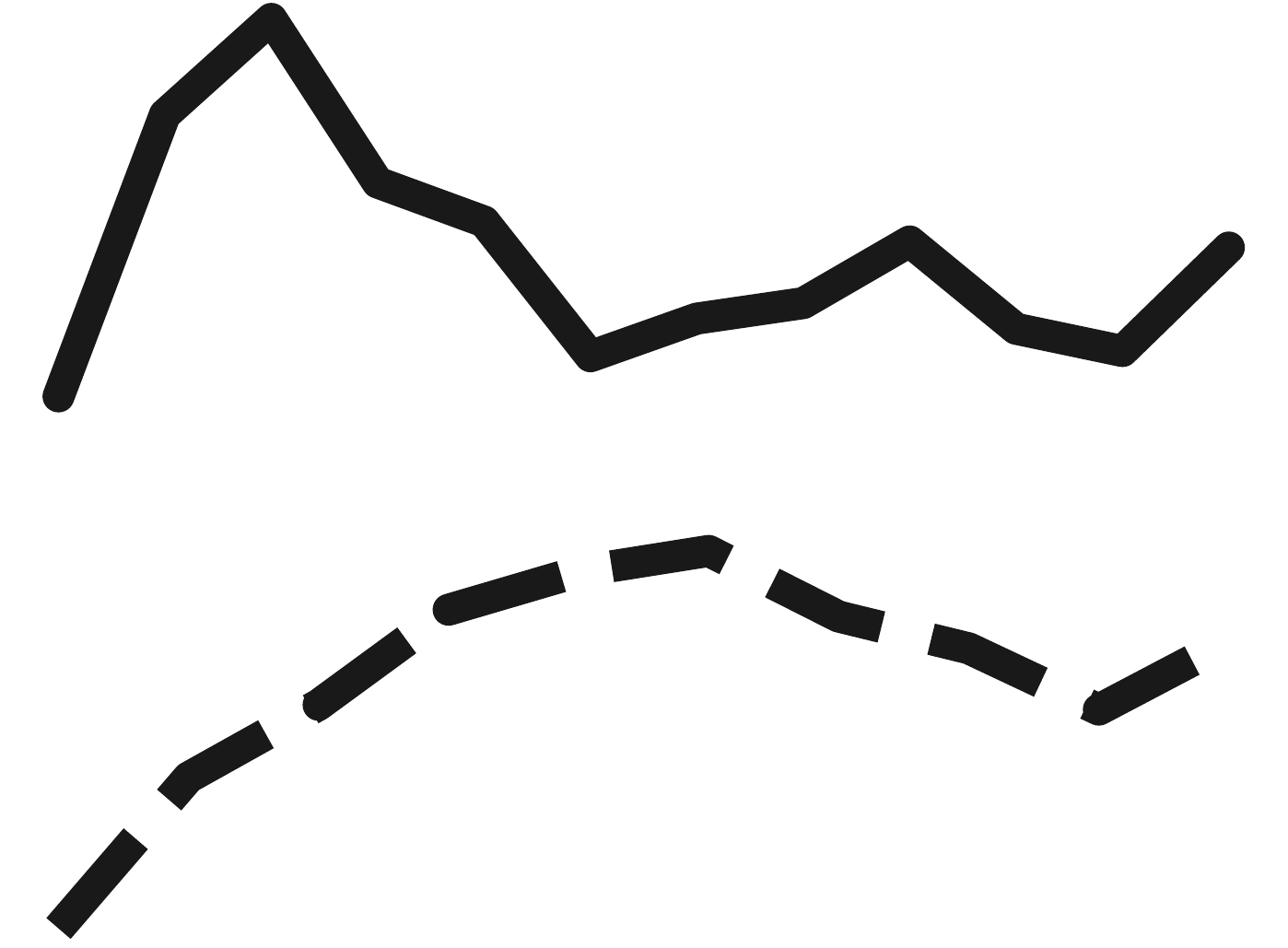}} \\
    \hline
    Resource Leaks (T8) & 10.6 (0.42) & \parbox[c]{0.7cm}{\includegraphics[width=0.8cm,keepaspectratio]{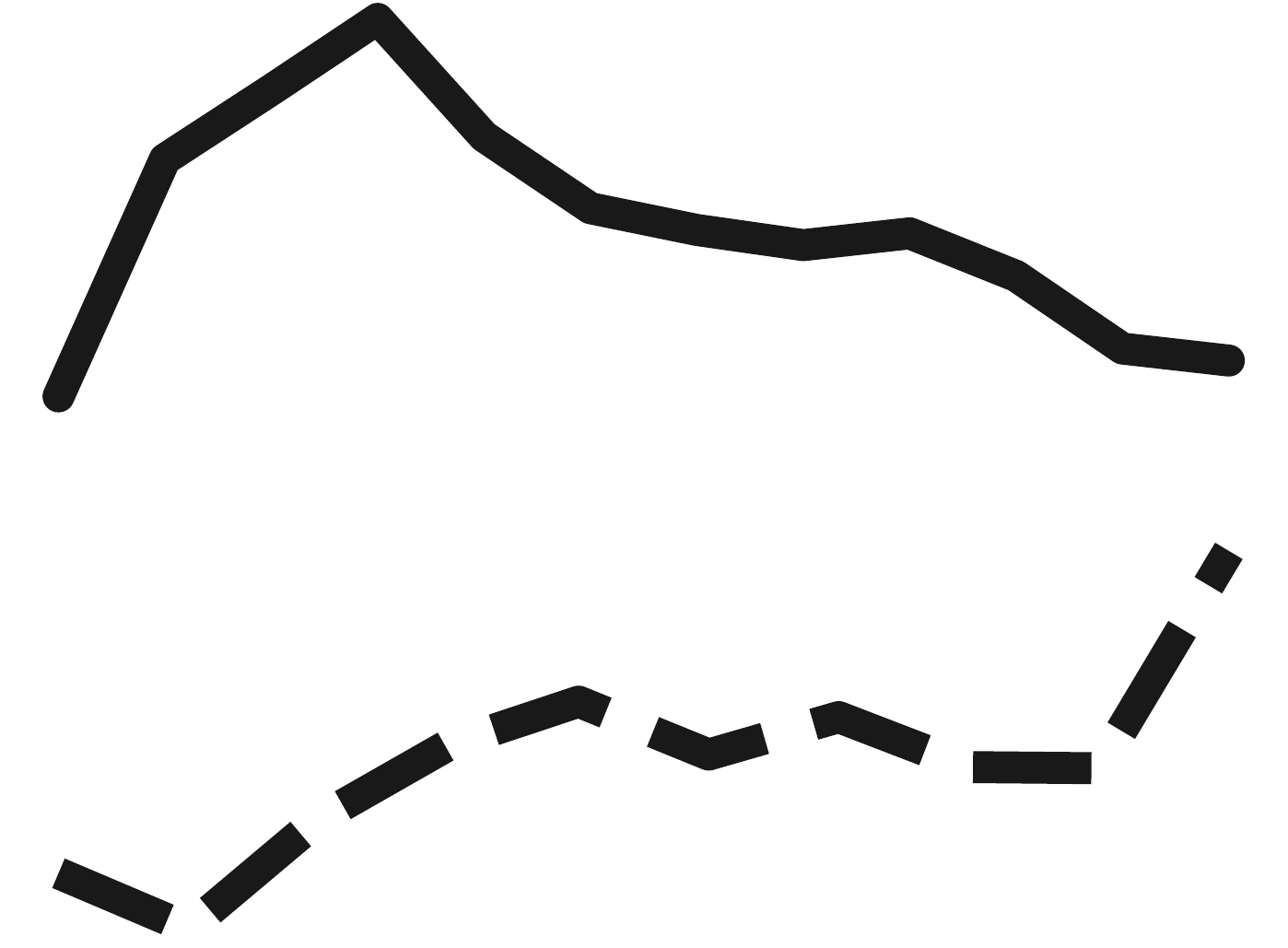}} \\
    \hline
    Network Attacks (T9) & 1.37 (8.79) & \parbox[c]{0.7cm}{\includegraphics[width=0.8cm,keepaspectratio]{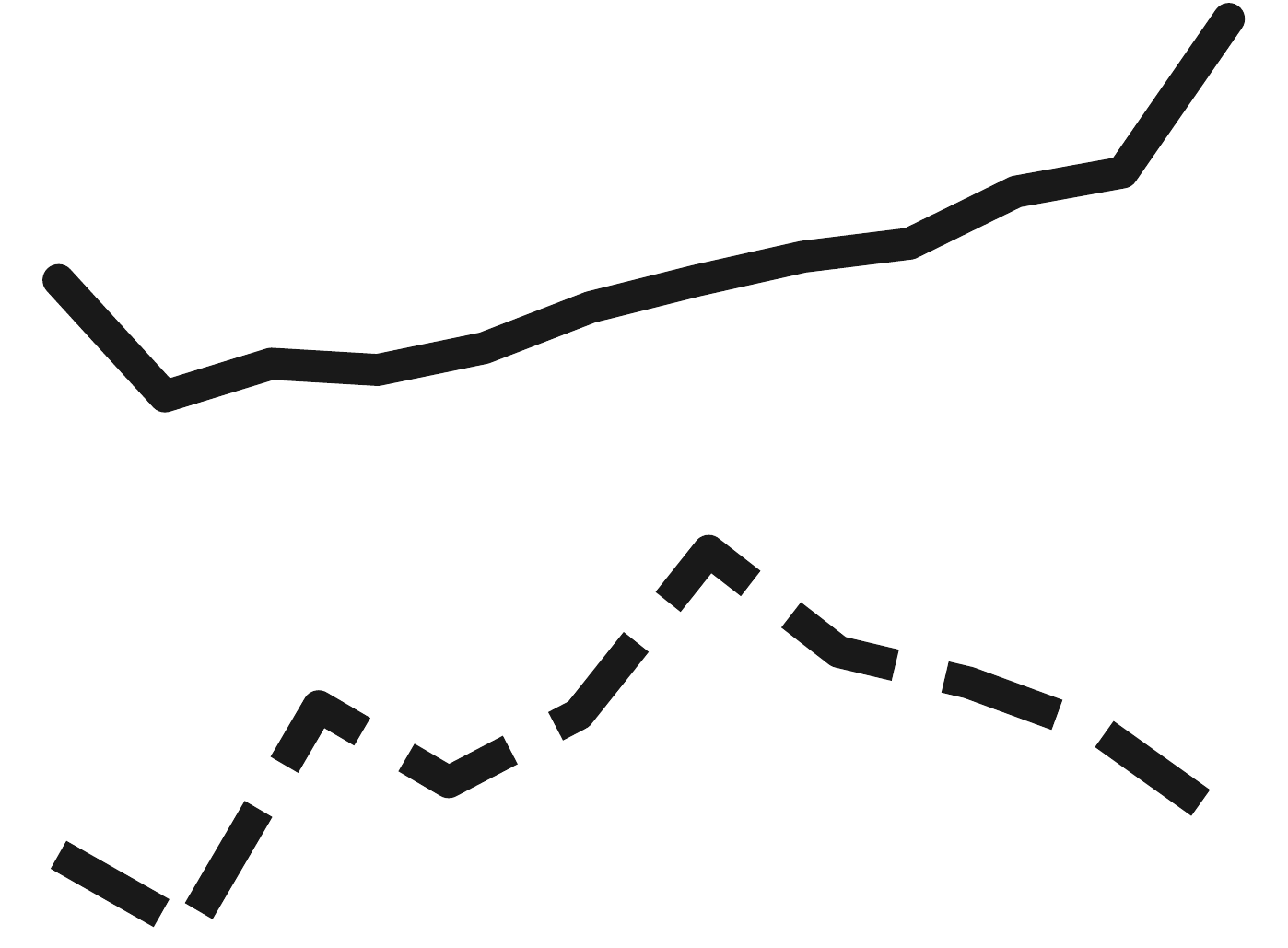}} \\
    \hline
    Memory Allocation Errors (T10) & \textbf{21.6} (2.82) & \parbox[c]{0.7cm}{\includegraphics[width=0.8cm,keepaspectratio]{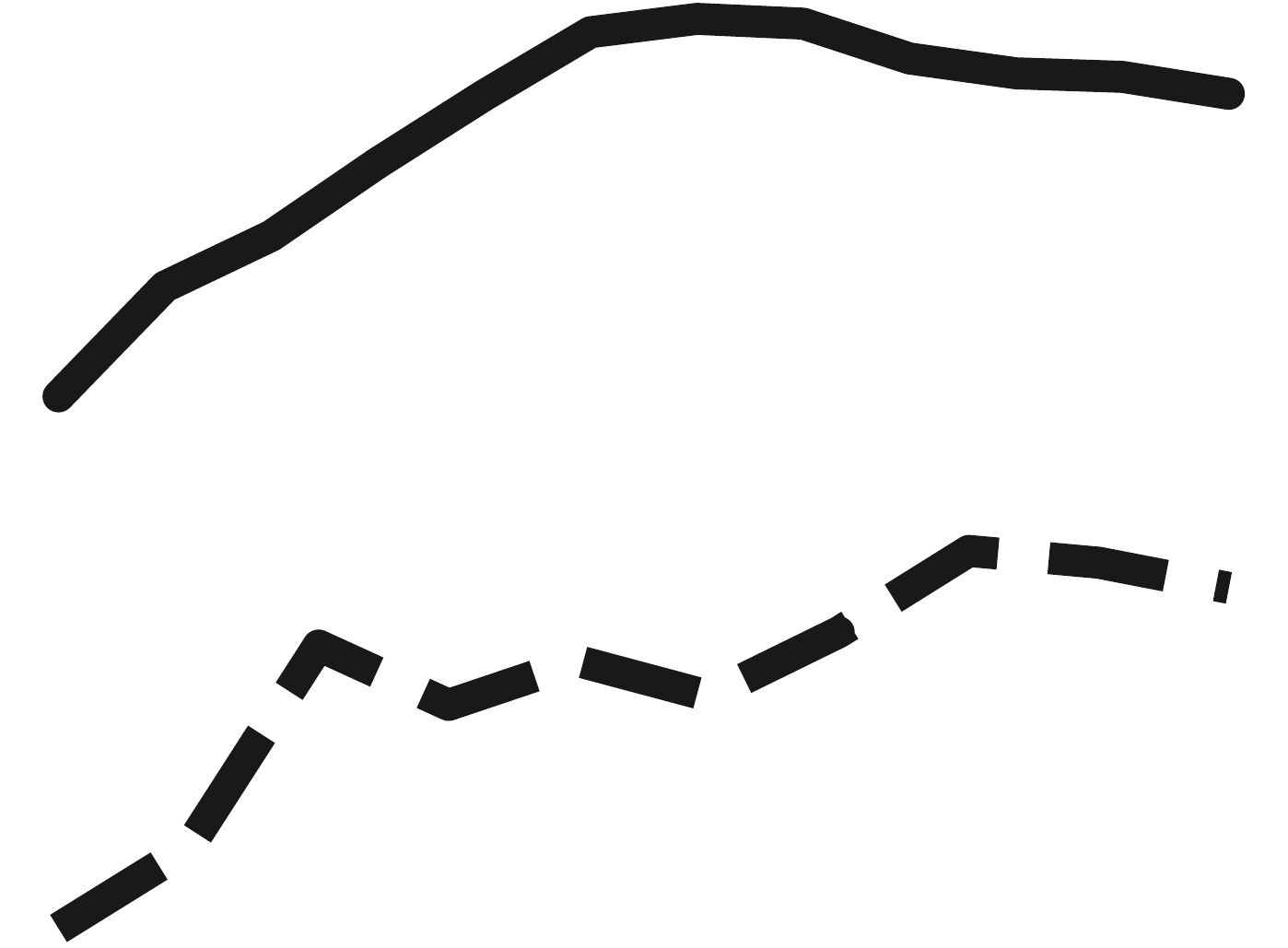}} \\
    \hline
    Cross-site Scripting (XSS) (T11) & 7.73 (8.09) & \parbox[c]{0.7cm}{\includegraphics[width=0.8cm,keepaspectratio]{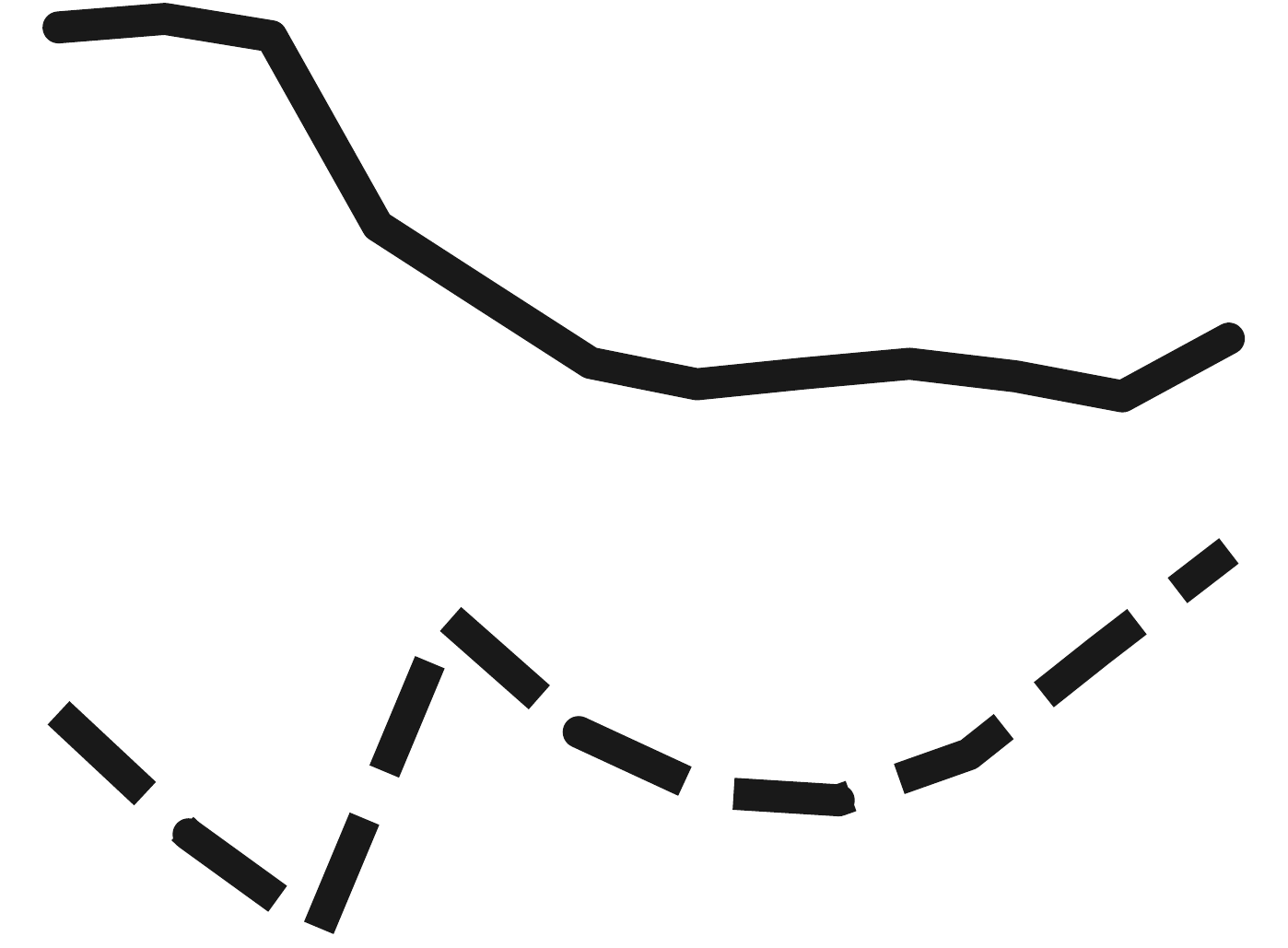}}  \\
    \hline
    Vulnerability Theory (T12) & 10.7 (\textbf{33.7}) & \parbox[c]{0.7cm}{\includegraphics[width=0.8cm,keepaspectratio]{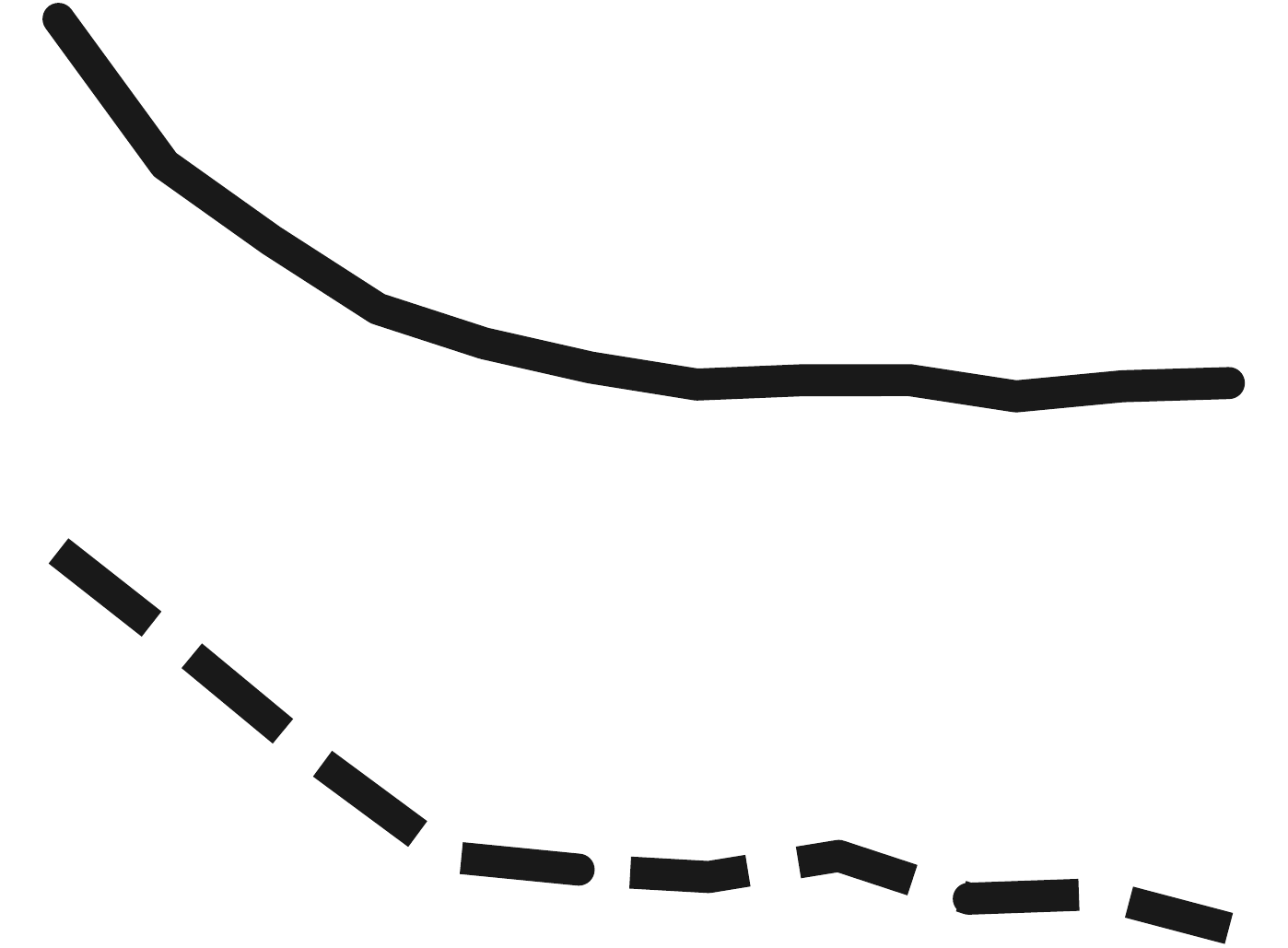}} \\
    \hline
    Brute-force/Timing Attacks (T13) & 1.08 (1.28) & \parbox[c]{0.7cm}{\includegraphics[width=0.8cm,keepaspectratio]{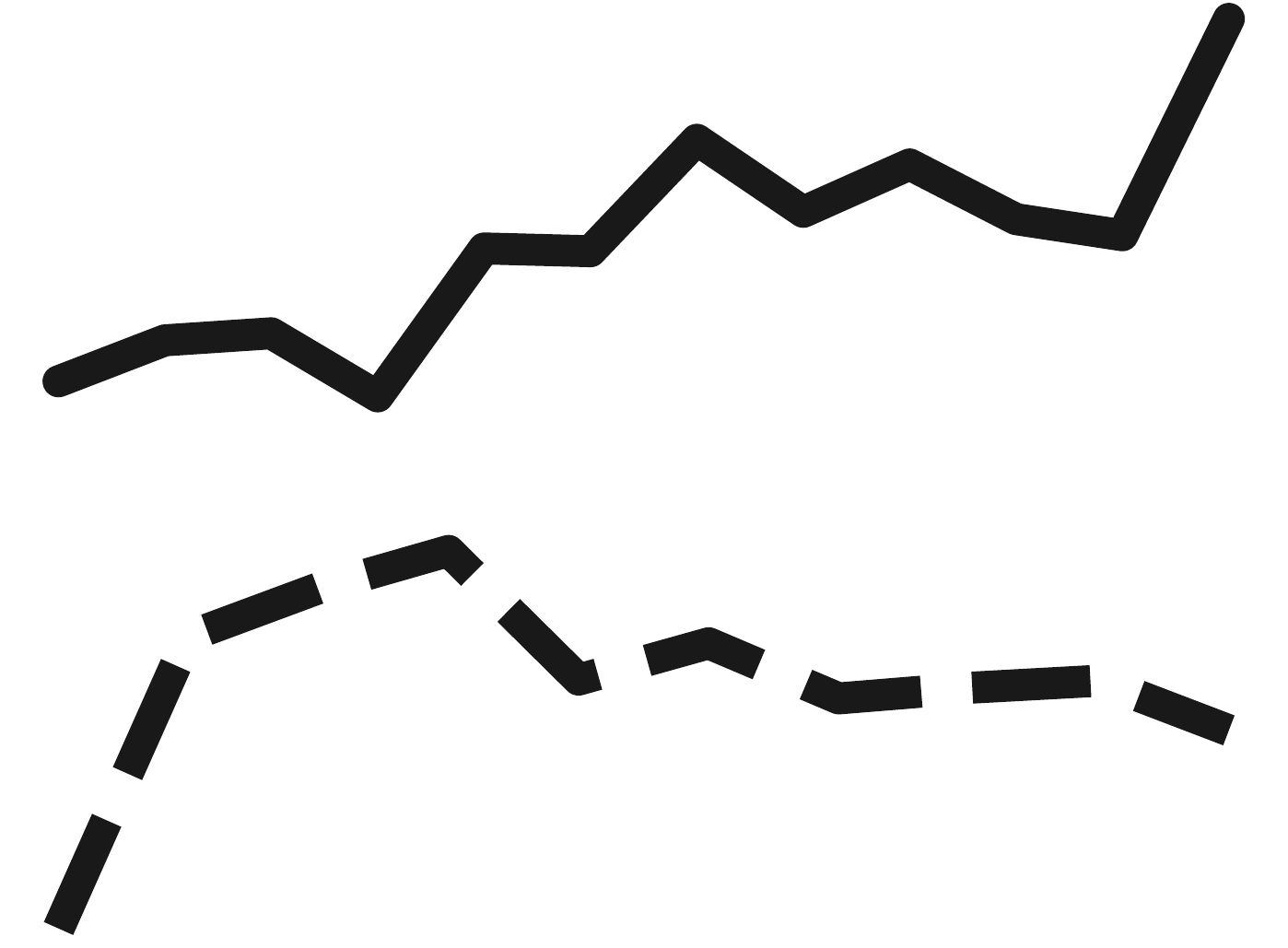}} \\
    \hline
    \end{tabular}%
  \label{tab:topic_list}%
  \vspace{-6pt}
\end{table}

\vspace{-10pt}

\section{Results}
\label{sec:results}

\subsection{\textbf{RQ1}: What are SV Discussion Topics on Q\&A Sites?}
\label{subsec:rq1_results}

Following the procedure in section~\ref{subsec:lda}, we identified \textit{13} SV topics (see Table~\ref{tab:topic_list}) on SO and SSE using the optimal LDA model with $\alpha=\beta=0.08$. We found LDA models having from 11 to 17 topics produced similar coherence metrics. Three of the authors manually examined these cases, as in~\cite{abdellatif2020challenges}. Duplicate and/or platform-specific topics (e.g., web and mobile) appeared from 14 topics, making the taxonomy less generalizable. 11 and 12 topics also had high-level topics (e.g., combining XSS and CSRF). Thus, 13 was chosen as the optimal number of SV topics. All the terms/posts of each SV topic can be found at~\cite{reproduction_package_ease2021}. We describe each topic hereafter with example SO/SSE posts. We examined 15 random posts per topic per site. If we identified some common patterns of discussions (e.g., attack vectors or assets) on a site, we would extract another 15 random posts of the respective site to confirm our observations. If a pattern was no longer evident in the latter 15 posts, we would not report it.

\textbf{Malwares (T1)}. This topic referred to the detection and removal of malicious code such as malwares, viruses and worms that infected devices, platforms and websites. T1 posts on SO were usually about malwares in content management systems such as Wordpress or Joomla (e.g., post 16397854: ``\textit{How to remove wp-stats malware in wordpress}'' or post 11464297: ``\textit{How to remove .htaccess virus}'). In contrast, SSE often discussed malwares/viruses coming from storage devices such as SSD (e.g., post 227115: ``\textit{Can viruses of one ssd transfer to another ssd?}'') or USB (e.g., post 173804: ``\textit{Can Windows 10 bootable USB drive get infected while trying to reinstall Windows?}'').

\textbf{SQL Injection (T2)}. This topic concerned tactics to properly sanitize malicious inputs that could modify SQL commands and pose threats (e.g., stealing or changing data) to databases in various programming languages (e.g., PHP, Java, C\#). A commonly discussed tactic was to use prepared statements, which also helped increase the efficiency of query processing. For example, developers asked questions like ``\textit{How to parameterize complex oledb queries?}'' (SO post 9650292) or ``\textit{How to make this code safe from SQL injection and use bind parameters}'' (SSE post 138385).

\textbf{Vulnerability Scanning Tools (T3)}. This topic was about issues related to tools for automated detection/assessment of potential SVs in an application. Discussions of T3 mentioned different tools, and OWASP ZAP was a commonly discussed one. For example, post 62570277 on SO discussed ``\textit{Jenkins-zap installation failed}'', while post 126851 on SSE asked ``\textit{How do I turn off automated testing in OWASP ZAP?}'' One possible explanation is that OWASP ZAP is a free and easy-to-use tool for detecting and assessing SVs that appear in the well-known top-10 OWASP list for web applications.

\textbf{Cross-site Request Forgery (CSRF) (T4)}. This topic contained discussions on proper setup and configuration of web application frameworks to prevent CSRF SVs. These SVs could be exploited to send requests to perform unauthorized actions from an end user that a web application trusts. Discussions covered various issues in implementing different CSRF prevention techniques recommended by OWASP.\footnote{\mbox{\label{fn:csrf_owasp}}https://cheatsheetseries.owasp.org/cheatsheets/Cross-Site\_Request\_Forgery\_Prevention\_Cheat\_Sheet.html}
Some commonly discussed techniques were anti-CSRF token (e.g., SO post 59664094: ``\textit{Why Laravel 4 CSRF token is not working?}''), double submit cookie (e.g., SSE post 203996: ``\textit{What is double submit cookie? And how it is used in the prevention of CSRF attack?}''), and SameSite cookie attribute (e.g., SO post 41841880: ``\textit{What is the benefit of blocking cookie for clicked link? (SameSite=strict)}'').

\textbf{File-related Vulnerabilities (T5)}. Discussions of this topic were about SVs in files that could be exploited to gain unauthorized access. The common SV types were Path/Directory Traversal via Symlink (e.g., SSE post 165860: ``\textit{Symlink file name - possible exploit?}''), XML External Entity (XXE) Injection (e.g., SO post 51860873: ``\textit{Is SAXParserFactory susceptible to XXE attacks?}''), and Unrestricted File Upload (e.g., SSE post 111935: ``\textit{Exploiting a PHP server with a .jpg file upload}''). These SVs usually occurred for Linux-based systems, suggesting that Linux is more popular for servers.

\textbf{Synchronization Errors (T6)}. This topic involved SVs produced through errors in synchronization logic (usually related to threads), which could slow down system performance. Some common SV types being discussed were deadlocks (e.g., SO post 38960765: ``\textit{How to avoid dead lock due to multiple oledb command for same table in ssis}'') and race conditions (e.g., SSE post 163209: ``\textit{What's the meaning of `the some sort of race condition' here?}'').

\textbf{Encryption Errors (T7)}. This topic included cryptographic issues leading to falsified authentication or retrieval of sensitive data, e.g., Man-in-the-middle (MITM) attack. Many posts discussed public/private keys for encryption/decryption, especially using SSL/TLS certificates to defend against MITM attacks (attempts to steal information sent between browsers and servers). Some example discussions are post 23406005 on SO (``\textit{Man In Middle Attack for HTTPS}'') or post 105773 on SSE (``\textit{How is it that SSL/TLS is so secure against password stealing?}''). This may imply that many developers are still not familiar with these certificates in practice.

\textbf{Resource Leaks (T8)}. This topic considered SVs arising from improper releases of unused memory which could deplete resources and decrease system performance. Many discussions of T8 were about memory leaks in mobile app development. Issues were usually related to Android (e.g., SO post 58180755: ``\textit{Deal with Activity Destroying and Memory leaks in Android}'') or IOS (e.g., SO post 47564784: ``\textit{iOS dismissing a view controller doesn't release memory}'').

\textbf{Network Attacks (T9)}. This topic discussed attacks carried out over an online computer network, e.g., Denial of Service (DoS) and IP/ARP Spoofing, and potential mitigations. These network attacks directly affected the availability of a system. For instance, SSE post 86440 discussed ``\textit{VPN protection against DDoS}'' or SO post 31659468 asked ``\textit{How to prevent ARP spoofing attack in college?}''.

\textbf{Memory Allocation Errors (T10)}. T10 and T8 were related to memory issues, but T10 did not consider memory release. Rather, this topic focused on SVs caused by accessing or using memory outside of what allocated that could be exploited to access restricted memory location or crash an application. In this topic, segmentation faults (e.g., SO post 31260018: ``\textit{Segmentation fault removal duplicate elements in unsorted linked list}'') and buffer overflows (e.g., SSE post 190714: ``\textit{buffer overflow 64 bit issue}'') were commonly discussed.

\textbf{Cross-site Scripting (XSS) (T11)}. This topic mentioned tactics to properly neutralize user inputs to a web page to prevent XSS attacks. These attacks could exploit users' trust in web servers/pages to trick them to execute malicious scripts and perform unwanted actions. XSS (T11) and CSRF (T4) are both client-side SVs, but XSS is more dangerous since it can bypass all countermeasures of T4.\mbox{\textsuperscript{\mbox{\ref{fn:csrf_owasp}}}} On SO and SSE, discussions covered all three types of XSS: (\textit{i}) reflected XSS (e.g., SSE post 57268: ``\textit{How does the anchor tag (<a>) let you do an Reflected XSS?}''), (\textit{ii}) stored/persistent XSS (e.g., SO post 54771897: ``\textit{How to defend against stored XSS inside a JSP attribute value in a form}''), and (\textit{iii}) DOM-based XSS (e.g., SO post 44673283: ``\textit{DOM XSS detection using javascript(source and sink detection)}'').

\textbf{Vulnerability Theory (T12)}. This topic focused on theoretical/social aspects and best practices in the SV life cycle. Many posts compared different SV-related terminologies, e.g., SSE post 103018 asked about ``\textit{In CIA triad of information security, what's the difference between confidentiality and availability?}'' or SO post 402936 discussed ``\textit{Bugs versus vulnerabilities?}''. Several other posts asked about internal SV reporting process (e.g., SO post 3018198: ``\textit{How best to present a security vulnerability to a web development team in your own company?}'') or public SV disclosure policy (e.g., SSE post: ``\textit{How to properly disclose a security vulnerability anonymously?}'').

\textbf{Brute-force/Timing Attacks (T13)}. T13 and T7 both exploited cryptographic flaws, but these two topics used different attack vectors/methods. T7 focused on MITM attacks, while T13 was about attacks making excessive attempts or capturing the timing of a process to gain unauthorized access. Some example posts of T13 are SO post 3009988 (``\textit{What's the big deal with brute force on hashes like MD5}'') or SSE post 9192 (``\textit{Timing attacks on password hashes}'').

\begin{table*}[t]
\fontsize{8}{9}\selectfont
  \centering
  \caption{General expertise in terms of average reputation of each topic on SO and SSE (in parentheses). \textbf{Notes}: The values were normalized by the max and min values of each category. T8 on SSE was excluded since it did not have any accepted answer.}
    \begin{tabular}{|l|c|c|c|c|c|c|c|c|c|c|c|c|c|}
    \hline
    \makecell{\textbf{General}\\ \textbf{expertise}} & \textbf{T1} & \textbf{T2} & \textbf{T3} & \textbf{T4} & \textbf{T5} & \textbf{T6} & \textbf{T7} & \textbf{T8} & \textbf{T9} & \textbf{T10} & \textbf{T11} & \textbf{T12} & \textbf{T13} \\
    \hline
    \textbf{Reputation} & \makecell{\textbf{0.00}\\ (0.16)} & \makecell{0.93\\ (0.05)} & \makecell{\textbf{0.01}\\ (\textbf{0.00})} & \makecell{0.28\\ (0.18)} & \makecell{0.43\\ (0.14)} & \makecell{0.88\\ (0.04)} & \makecell{0.51\\ (0.72)} & \makecell{0.49\\ (--)} & \makecell{0.07\\ (0.24)} & \makecell{0.62\\ (0.34)} & \makecell{0.84\\ (0.22)} & \makecell{0.59\\ (0.29)} & \makecell{\textbf{1.00}\\ (\textbf{1.00})} \\
    \hline
    \end{tabular}%
  \label{tab:gen_exp}%
\end{table*}

\textbf{Proportion and Evolution of SV Topics}. We analyzed the proportion (share metric in Eq.~\eqref{eq:share}) and the evolution trend of SV topics from their inception on SO (2008) and SSE (2010) to 2020 (see Table~\ref{tab:topic_list}). The topic patterns and dynamics of SO were different from those of SSE. Specifically, Memory Allocation Errors (T10) had the greatest number of posts on SO, while Vulnerability Theory (T12) had the largest proportion on SSE. Apart from XSS (T11) and Brute-force/Timing Attacks (T13), topics with many posts in one source were not common in the other source. Moreover, we discovered three consistent topic trends on both SO and SSE: Malwares (T1) ($\nearrow$), CSRF (T4) ($\nearrow$), File-related SVs (T5) ($\nearrow$) and Vulnerability Theory (T12) ($\searrow$). Among them, CSRF had the fastest changing pace.
These trends were confirmed significant with p-values $<$ 0.05 using Mann-Kendall non-parametric trend test~\cite{mann1945nonparametric}.

\vspace{-10pt}

\subsection{\textbf{RQ2}: What are the Popular and Difficult SV Topics on Q\&A Sites?}
\label{subsec:rq2_results}

\begin{figure}[t]
    \centering
    {%
    \setlength{\fboxsep}{0pt}%
    \setlength{\fboxrule}{0.5pt}%
    \fbox{\includegraphics[width=0.98\columnwidth,keepaspectratio]{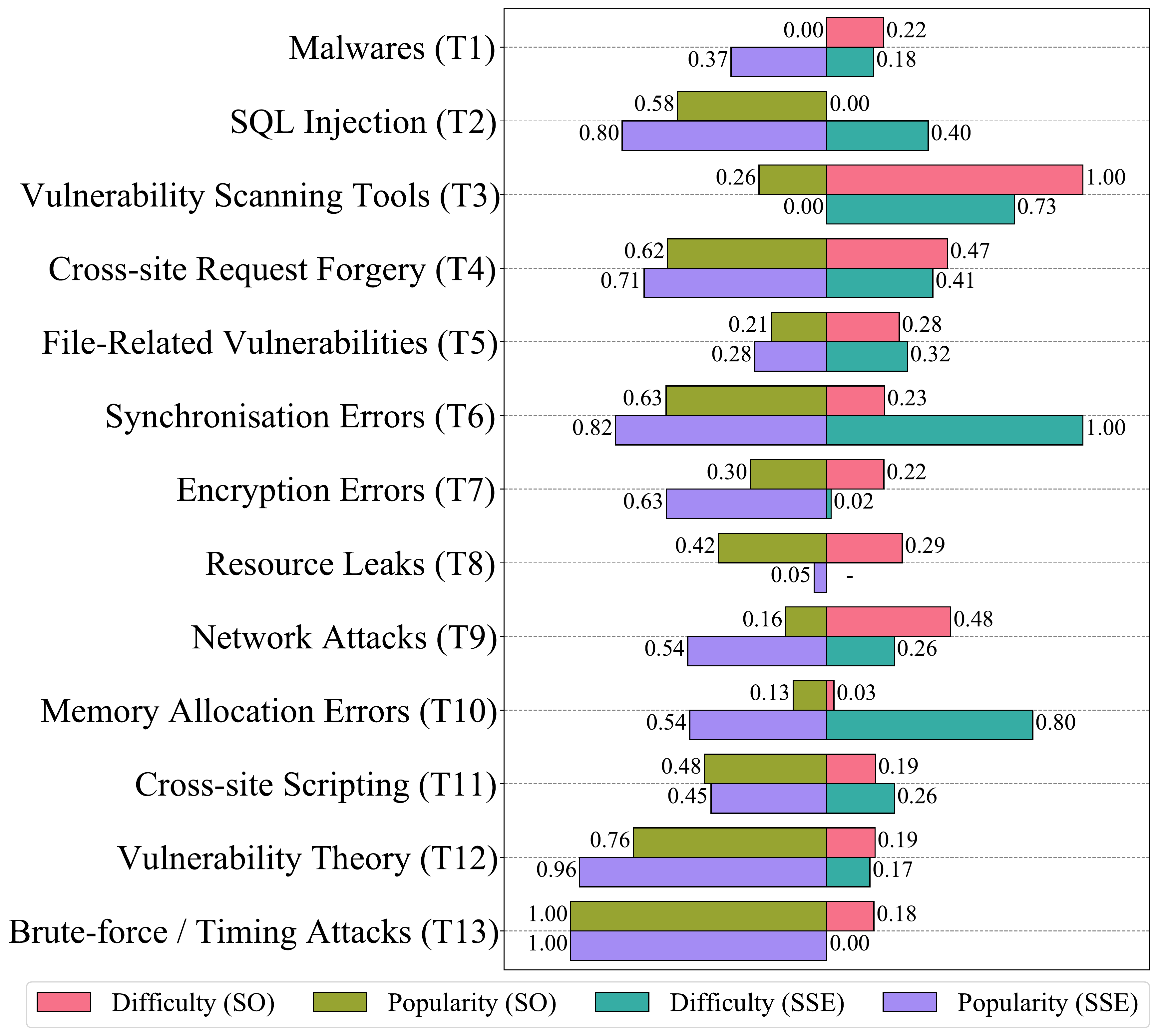}
    }}
    \caption{Popularity and difficulty of 13 SV topics on SO and SSE. \textbf{Notes}: The values were normalized by the max and min values of each category. Difficulty of T8 on SSE was excluded since it did not have any accepted answer.}
    \label{fig:pop_diff}
\end{figure}

As shown in Fig.~\ref{fig:pop_diff}, the popularity and difficulty of 13 identified SV topics were different between SO and SSE. For conciseness, we only report the geometric means of the popularity and difficulty metrics in this section. The complete values of individual metrics (see section~\mbox{\ref{rq2_method}}) can be found at~\mbox{\cite{reproduction_package_ease2021}}.

\textbf{Topic Popularity}. Brute-force/Timing attacks (T13) and Vulnerability Theory (T12) were the top-2 most popular topics. Despite being the most popular topic, T13 only had 1.1\% and 1.3\% posts on SO and SSE, respectively. Conversely, Memory Allocation Errors (T10) had the most posts on SO (RQ2), but T10 was only the second least popular topic. We found no significant correlation between the topic popularity and share metric with Kendall's Tau correlation test~\cite{knight1966computer} at 95\% confidence level. These findings suggest that share metric does not necessarily reflect the topic popularity since it does not consider user's activities on Q\&A sites.

\textbf{Topic Difficulty}. The most difficult topics were not popular or associated with many posts, i.e., Vulnerability Scanning Tools (T3) and Network Attacks (T9) on SO as well as T3, Synchronization Errors (T6) and Memory Allocation Errors (T10) on SSE. The high difficulty of T3 on both sites was potentially caused by the low familiarity with a wide array of vendors and tools available for SV detection and assessment~\cite{kritikos2019survey}. Some topics with many posts (high share metric) like Memory Allocation Errors (T10) and SQL Injection (T2) were the two easiest ones on SO despite being significantly more difficult on SSE. On the contrary, Malwares (T1) and Network Attacks (T9) were more popular yet easier on SSE. These numbers suggest that it may be better to ask the topics T2, T8 (only a few posts on SSE) and T10 on SO to obtain answers faster, while asking T1 and T9 on SSE would be more optimal. However, the topic difficulty did not correlate with either the topic popularity or share metric on both SO and SSE, confirmed using Kendall's Tau test~\cite{knight1966computer} with a confidence level of 95\%. With the same confidence level, no significant differences in terms of average topic-wise popularity and difficulty between SO and SSE were recorded using non-parametric Mann-Whitney U-test~\cite{mann1947test}.

\subsection{\textbf{RQ3}: What is the Level of Expertise to Answer SV Questions on Q\&A Sites?}
\label{subsec:rq3_results}

\textbf{General Expertise}. The average general expertise (reputation) of the accepted answerers in SV posts was 1.3 to 5.8 times higher than those of generic posts~\cite{barua2014developers}, general security~\cite{yang2016security}, mobile development~\cite{rosen2016mobile}, concurrency~\cite{ahmed2018concurrency}, machine learning~\cite{bangash2019developers} and deep learning~\cite{han2020programmers}. The higher reputation values were confirmed with p-values $<$ 0.05 (significance level) using non-parametric Mann-Whitney U-test~\mbox{\cite{mann1947test}}.
However, the average percentage of the same users who got accepted answers on both SO and SSE was quite small across topics, i.e., 1\% to 18\%, implying not much SV knowledge sharing between the two sites. The average topic-wise reputation on SO was higher than that of SSE with a p-value of 0.001 (Mann-Whitney U-test). This might be because SO users could engage in many more posts of different topics (not only security). Table~\ref{tab:gen_exp} reports the general expertise of 13 SV topics. On SO, Brute-force/Timing Attacks (T13), SQL Injection (T2), Synchronization Errors (T6) and XSS (T11) were the topics that experts focused on the most. On SSE, T13 again and Encryption Errors (T7) were the topics of interest for experts. In contrast, Malwares (T1), Vulnerability Scanning Tools (T3) and Network Attacks (T9) on SO did not attract as much attention from experts.
On SSE, T3 was also of the least interest to experts. Overall, experts on Q\&A sites tended to favor the SV topics with high popularity and low difficulty, confirmed with p-values $<$ 0.05 using Kendall's Tau correlation test~\cite{knight1966computer}.

\begin{figure}[t]
    \centering
    {%
    \setlength{\fboxsep}{0pt}%
    \setlength{\fboxrule}{0.5pt}%
    \fbox{\includegraphics[width=0.98\columnwidth,keepaspectratio]{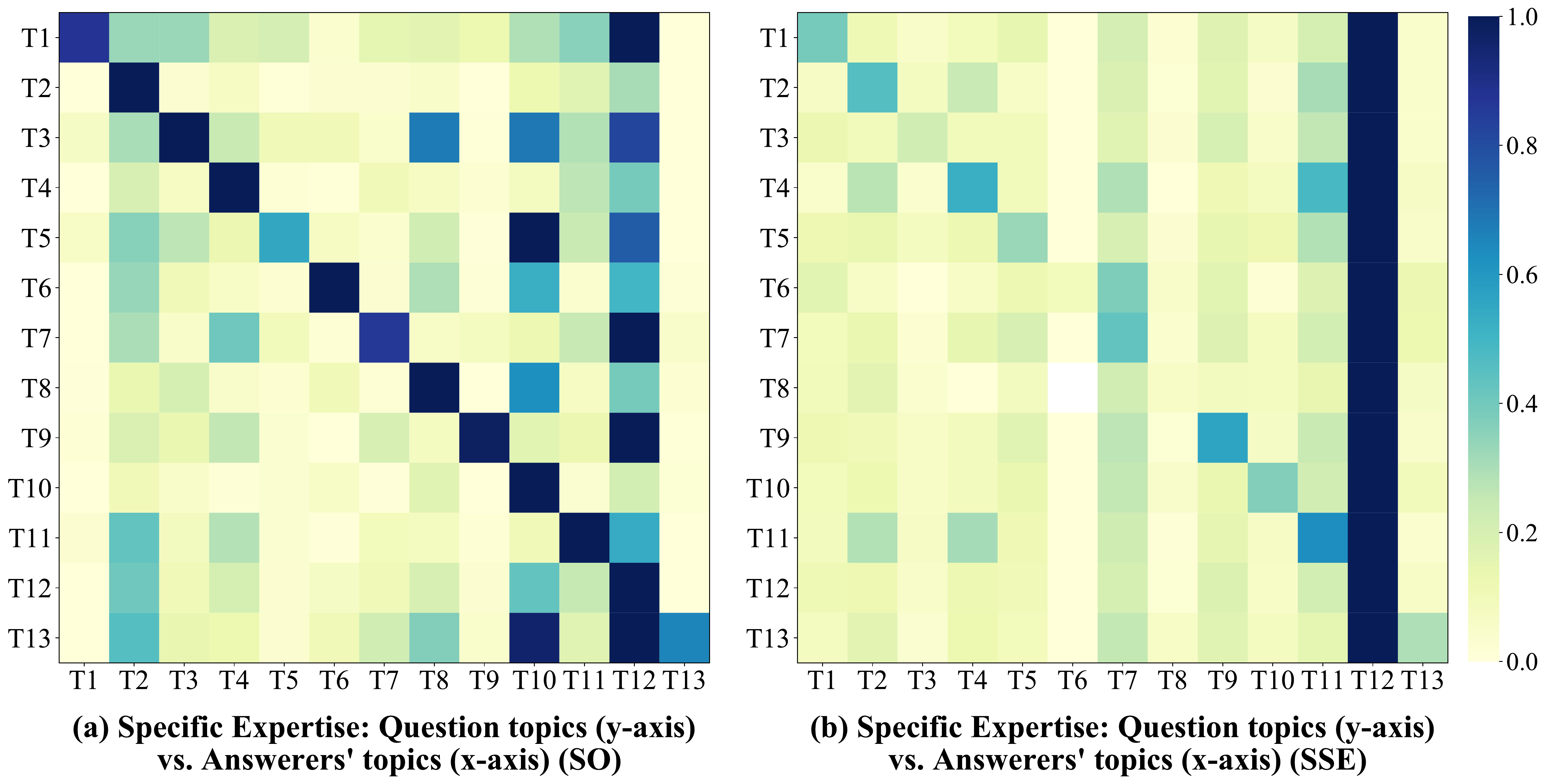}
    }}
    \caption{Topic correlations between SV questions \& answerers' SV specific knowledge on SO (a) \& SSE (b). \textbf{Notes}: Light to dark color shows weak to strong correlation. Each cell was normalized by max and min values of each question topic.}
    \label{fig:qa_topic_correlation}
\end{figure}

\textbf{Specific Expertise}.
Fig.~\mbox{\ref{fig:qa_topic_correlation}} shows the correlation between the pairs of question SV topics and answerers' SV topics (see Eq.~\mbox{\eqref{eq:spec_exp}}). The most frequent answerers' SV topic was Vulnerability Theory (T12). On SSE, frequent answerers for T12 could answer every question topic. On SO, besides T12, users specialized in Memory-related Errors (T8 and T10) also answered the questions of other SV topics. These patterns might be because of the prevalence (RQ2) of topics T8 and T10 on SO as well as T12 on SSE. Conversely, Malwares (T1), Network Attacks (T9) and Brute-force/Timing Attacks (T13) on SO as well as Synchronization Errors (T6), Resource Leaks (T8) and T13 on SSE had unique answerers (i.e., users who usually answered questions of only one topic in the SV domain).
Furthermore, on SO, most answerers were relevant for each SV topic (dark color on the diagonal in Fig.~\mbox{\ref{fig:qa_topic_correlation}}a), but it was not always the case on SSE (see Fig.~\mbox{\ref{fig:qa_topic_correlation}}b). Such results suggest that it may be easier to find relevant answerers for different SV topics on SO than on SSE.

\subsection{\textbf{RQ4}: What Types of Answers are Given to SV Questions on Q\&A Sites?}
\label{subsec:rq4_results}

\begin{table*}[t]
\fontsize{8}{9}\selectfont
  \centering
  \caption{Answer types of SV discussions identified on Q\&A websites. \textbf{Note}: An answer can have more than one solution type.}
  \vspace{-6pt}
    \begin{tabular}{|p{0.13\textwidth}|p{0.4\textwidth}|p{0.18\textwidth}|p{0.185\textwidth}|}
    \hline
    \makecell[l]{\textbf{Answer type of} \\ \textbf{SV discussions}} & \multicolumn{1}{c|}{\textbf{Description \& Example Posts}} & \makecell[c]{\textbf{Top-3 related}\\ \textbf{question types~\cite{treude2011programmers}}} & \makecell[c]{\textbf{Proportion (\%) on SO}\\ \textbf{\& SSE (in parenthesis)}} \\
    \hline
    (Dis-)Confirmation (DC/Co) & Confirm/agree or refute/disagree with a major point or concept made by the asker (e.g., SO post 16155188 or SSE post 31306) & Decision Help, How-to, Conceptual & \makecell[c]{11.5 (23.7)} \\
    \hline
    Explanation (Ex) & Explain concepts, definitions and ``why'' to take certain actions (e.g., SO post 53446941 or SSE post 157240) & Decision Help, Conceptual, Discrepancy & \makecell[c]{14.6 (27.1)} \\
    \hline
    Error (Er) & Point out an error in the source code or another attachment of the initial question (e.g., SO post 29750534 or SSE post 159907) & Discrepancy, Error, How-to & \makecell[c]{13.0 (2.3)} \\
    \hline
    Action to Take (AT) & Describe step/action(s) (``how-to'') to solve a problem (e.g., SO post 22860382 or SSE post 180053) & How-to, Discrepancy, Decision Help & \makecell[c]{\textbf{22.5} (15.4)} \\
    \hline
    External Source (ES) & Provide reference/link to external source(s) (e.g., SO post 445177 or SSE post 107498) & Decision Help, How-to, Discrepancy & \makecell[c]{18.1 (\textbf{28.7})} \\
    \hline
    Code Sample (CS) & Provide an explicit example of code snippet (e.g., SO post 20763476 or SSE post 36804) & Discrepancy, How-to, Error & \makecell[c]{16.6 (2.0)} \\
    \hline
    Self-Answer (SA) & Answer given by the same user who submitted the question (e.g., SO post 55784402 or SSE post 100761) & Discrepancy, Error, How-to & \makecell[c]{3.7 (1.0)} \\
    \hline
    \end{tabular}%
  \label{tab:answer_types}%
\end{table*}

\begin{table}[t]
\fontsize{8}{9}\selectfont
  \centering
  \caption{Top-1 answer types of 13 SV topics on SO \& SSE (in parenthesis). Note: T8 on SSE was excluded since it did not have any accepted answer.}
  \vspace{-6pt}
    \begin{tabular}{|c|c||c|c|}
    \hline
    \textbf{Topic} & \textbf{Top-1 Answer Type} & \textbf{Topic} & \textbf{Top-1 Answer Type}\\
    \hline
    T1 & AT/ES (Ex) & T8 & ES (--) \\
    \hline
    T2 & CS (ES) & T9 & DC/Co/ES/SA (Ex) \\
    \hline
    T3 & ES (ES) & T10 & AT (Ex)  \\
    \hline
    T4 & ES (DC/Co) & T11 & AT (DC/Co) \\
    \hline
    T5 & Ex (ES) & T12 & Ex (ES)  \\
    \hline
    T6 & Ex/ES (DC/Co/ES) & T13 & Ex/ES (Ex) \\
    \hline
    T7 & Ex/ES (Ex) & -- & -- \\
    \hline
    \end{tabular}%
  \label{tab:answers_per_topic}%
  \vspace{-6pt}
\end{table}

Our open coding process in RQ4 identified \textit{seven} answer categories of SV discussions on Q\&A sites, as shown in Table~\ref{tab:answer_types}. Some answer types provided experience (DC/Co and Er) or language/platform-specific support (AT, ES and CS), which is hardly found on expert-based security sites (e.g., CWE or OWASP). We correlated such answer types with the question categories of Treude et al.~\cite{treude2011programmers}.
We found reasonable matches between answer and question types, e.g., (dis)agreeing (DC/Co) with a decision (Decision Help), explaining (Ex) a concept (Conceptual), and providing different solutions to resolve unexpected situations (Discrepancy). Discrepancy and Error were also among the most frequent question types, supporting that our posts were about issues/errors in addressing SVs.

\textbf{Site-wise answer types}. According to Table~\ref{tab:answer_types}, Action to Take (AT) and External Sources (ES) were the most common answer types on SO and SSE, respectively; whereas, Self-Answer (SA) was the least frequent one on both sites. We also noticed that both sites usually referred to external sources (ES). The most common sources included Wikipedia, other posts (e.g., related answers), GitHub issues/commits, product documentation (e.g., PHP, MySQL and Android) and SV sources (e.g., CVE (Common Vulnerabilities and Exposures), NVD, CWE, OWASP, CVE Details\footnote{www.cvedetails.com} and Exploit-DB\footnote{www.exploit-db.com}). Note that some provided links were unavailable or no longer maintained (e.g., CVE Details). Overall, the answers to SV questions on SO frequently provided detailed instructions (AT) and/or code samples (CS), while SSE tended to share more experience (DC/Co and Ex) to help.

\textbf{Topic-wise answer types}. We extend the site-wise findings to individual topics to enable developers to select the respective site (SO vs. SSE) based on their preferable SV solution types, as shown in  Table~\mbox{\ref{tab:answers_per_topic}}. Specifically, SO highlighted steps (AT) to fix Malwares (T1), Memory Allocation Errors (T10) and XSS (T11) as well as provided code snippets for SQL Injection (T2). On the other hand, SSE gave more relevant sources and explanations for such topics. One may argue that these different answer types were because of the different question types between SO and SSE, but we did not find any such significant differences for these topics. Instead, the fact that SO and SSE had quite different accepted answerers, as shown in RQ3, probably led to such different solution types.
There were four topics, namely SV Scanning Tools (T3), Synchronization \& Encryption Errors (T6 \& T7), Brute-force/Timing Attacks (T13), sharing similar top solutions on both SO and SSE. The remaining topics (T4, T5, T8 and T9) were mostly answered with explanation (Ex) or external sources (ES) on both SO and SSE.

\section{Discussion}
\label{sec:discussion}

\subsection{SV Discussion Topics on Q\&A Sites vs. Existing Security Taxonomies} \label{subsec:vs_literature}

\textbf{SV-specific topics and their support on Q\&A sites}. Compared to Yang et al.'s taxonomy~\mbox{\cite{yang2016security}}, we found related topics: T2, T3, T5, T10, T11 and T13, but we still had the following important differences. Firstly, our topics were emphasized more on security flaws, e.g., issues with encryption/decryption algorithms (T7) than how to implement/use them as in~\mbox{\cite{yang2016security}}. Secondly, we identified SV-specific topics previously unreported in~\mbox{\cite{yang2016security}}: Malwares (T1), CSRF (T4), Synchronization Errors (T6), Resource Leaks (T8), Network Attacks (T9) and Vulnerability Theory (T12). These SV topics show the necessity of focusing on SV-specific posts instead of general security ones. Thirdly, unlike~\mbox{\cite{yang2016security}}, we did not consider language-dependent topics (i.e., PHP, Flash, Javascript, Java and ASP.NET), helping our topics be more generalizable (e.g., XSS can occur in both PHP and ASP.NET). Mansooreh et al.~\mbox{\cite{zahedi2018empirical}} also devised a security taxonomy for GitHub issues; however, they focused on security features and implementation instead of any specific SV types. Note that we studied SV posts on both SO and SSE, while the existing studies only used one source of data (SO), enhancing the generalizability of our study. Specifically, we shed light on the differences between SV discussion topics on SO and those on SSE in terms of their proportions (RQ1), popularity/difficulty (RQ2), level of expertise (RQ3) and types of answers (RQ4). Our findings can be leveraged to select suitable site (i.e., more popular/experts, less difficult or having certain answer types) for asking different SV questions.

\textbf{Disconnection between SV discussions and expert-based SV sources}. Two authors manually mapped 13 SV topics with CWEs.\footnote{Due to limited space, we put our CWE mappings in the reproduction package~\mbox{\cite{reproduction_package_ease2021}.}} The agreement between the two authors was strong (Kappa score~\mbox{\cite{mchugh2012interrater}} was 0.892), and disagreements were resolved during a discussion section with the third author. We found that only seven of them were overlapping with the two well-known expert-based SV taxonomies: top-25 CWE and top-10 OWASP.\footnote{https://cwe.mitre.org/top25/ \& https://owasp.org/www-project-top-ten/} The overlapping topics were T2 (SQL Injection), T4 (CSRF), T5 (i.e., Path-traversal and Unrestricted File Upload), T7 (i.e., Improper Certificate Validation), T8 (Resource Leaks), T10 (Memory Allocation Errors) and T11 (XSS). There was no CWE for T3 and T12 since they mainly discussed SV scanning tools and/or socio-technical issues, respectively. Using keyword matching, we further found that only 159 and 71 out of a total of 839 CWEs were mentioned on SO and SSE, respectively; and only 20 and two CWEs appeared more than 100 times on SO and SSE, respectively.
Moreover, the fast increase of CSRF (T2) in RQ1 is noteworthy given that this SV type has been removed from the top-10 OWASP since 2013. We observed that many developers were aware of CSRF prevention techniques, but it was not always easy to apply these techniques and/or use built-in CSRF protection of a web framework (e.g., Spring Security) in practice. These results imply that expert-based sources do not always provide details on how to use/configure and implement/debug the reported SV prevention measures/tools in different use cases. The observed strong disconnection in the SV patterns between expert-based sources and discussions on Q\&A sites strengthens our motivation to study developers' real-life concerns in addressing SVs.

\subsection{Implications of Our Study}

\noindent \textbf{Researchers}. Different types of SVs are commonly discussed on Q\&A sites (RQ1), especially the prevalent, popular and increasing ones like Brute-force/Timing Attacks, Memory/File-related SVs, Malwares and CSRF. Researchers should develop robust detection, assessment and fixing methods for these SV types. Moreover, abundant off-the-shelf testing/scanning tools with complex configurations and different versions have made Vulnerability Scanning Tools one of the most difficult SV topics on Q\&A sites (RQ2). This motivates in-depth comparative study to investigate the effectiveness of available tools in different scenarios, especially for the SV types reported in this work. Further, researchers can develop techniques to search relevant users from multiple developer Q\&A sites (e.g., SO and SSE) rather than just SO to increase the cross-site knowledge sharing for difficult SV topics with low level of expertise (RQ3).

\noindent \textbf{Practitioners}.
Many of the SV topics identified in RQ1 have been induced by malicious inputs (i.e., SQL Injection, File-related SVs, CSRF and XSS). Hence, checking for proper input validation and neutralization should be top priorities in security testing. We have also observed that OWASP ZAP has been commonly used to support automated testing for these SV types. However, understanding, using and integrating SV scanning tools are still quite challenging (RQ2/RQ3). This suggests that tool developers/maintainers should further improve the functionalities, documentation and tutorials of their tools, potentially by leveraging SV-related questions/answers on Q\&A sites.
Moreover, practitioners should not expect to find information about zero-day SVs on SO and SSE. Instead, they can discuss how to identify/fix known SVs on SO and ask theoretical/social questions about SVs on SSE (RQ4). Some links in answers have been also found obsolete in RQ4; thus, practitioners should test and not always trust given solutions on Q\&A sites.

\vspace{-10pt}

\subsection{Threats to Validity}

Our data collection is the first threat. We might have missed some SV posts, but we followed standard techniques in the literature. It is hard to guarantee 100\% relevance of the retrieved posts without exhaustive manual validation, which is nearly impossible with more than 70k posts. However, this threat was greatly reduced since the selected posts were carefully checked by three of the authors.

The identified taxonomies can be another concern. Topic modeling with LDA has been shown effective for processing large amount of textual posts, but there is still subjectivity in labeling the topics. We mitigated this threat by manually examining at least 30 posts per topic and cross-checking with three of the authors. We also performed a similar manual checking for the answer types in RQ4.

The generalizability of our study may be a threat as well. The patterns we found may not be the same for other Q\&A sites and domains. However, the reported patterns for SV discussions on SO and SSE were at least confirmed significant using statistical tests with p-values $<$ 0.05. We also released our code and data at~\cite{reproduction_package_ease2021} for replication and extension to other domains.


\section{Conclusions and Future Work}
\label{sec:conclusions}

Through a large-scale study of 71,329 posts on SO and SSE, we have revealed the support of SV-focused discussions on Q\&A sites.
Using LDA, we devised 13 commonly discussed SV topics on Q\&A sites. Among these topics, we discovered the popular (e.g., Brute-force/Timing Attacks) and difficult (e.g., Vulnerability Scanning Tools) ones. The expertise for SV topics was high, but the knowledge sharing between the sites was still limited, and some topics (e.g., Vulnerability Scanning Tools) required more attention from experts. We also identified seven answer types for SV questions, in which SO offered more code-based/step-by-step solutions, while SSE provided more explanatory/experience-based replies. Overall, Q\&A sites do support SV discussions, but there is still a fair disconnection between SO and SSE. More effort is required to motivate cross-site engagement to better support (difficult) SV topics.

In the future, we aim to investigate SV topics on other Q\&A sites. We also plan to correlate the findings of SV discussions on Q\&A sites with SV detection and fixing activities on version control systems such as GitHub.



\vspace{-5pt}

\begin{acks}
The work was supported by the Cyber Security Research Centre Limited whose activities are partially funded by the Australian Government's Cooperative Research Centres Programme.
\end{acks}

\vspace{-5pt}

\bibliographystyle{ACM-Reference-Format}
\bibliography{reference}

\end{document}